\documentclass[12pt]{article}
\usepackage{amsfonts,amssymb,epsfig,amsmath,mathtools,xcolor}
\usepackage{color}

\usepackage{booktabs}

\usepackage{mathrsfs}
\usepackage{graphicx}
\usepackage{subcaption}
\usepackage{hyperref}
\usepackage{float}
\usepackage{braket}

\renewcommand{\baselinestretch}{1.2}

\newcommand{\nn}{\nonumber}

\newcommand{\RR}{\mathbb{R}}

\begin{document}

\makeatletter \@addtoreset{equation}{section} \makeatother
\renewcommand{\theequation}{\thesection.\arabic{equation}}
\renewcommand{\thefootnote}{\alph{footnote}}

\begin{titlepage}

\begin{center}
\hfill %{\tt SNUTP26-nnn}
\\

\vspace{2cm}

{\Large\bf Spectral function, OPE, and QNMs detect different singular points of black branes with unequal R-charges}

\vspace{2cm}

\renewcommand{\thefootnote}{\alph{footnote}}

{\large Hyunwoo Oh}

\vspace{0.7cm}

\textit{Department of Physics and Astronomy \& Center for
Theoretical Physics,\\
Seoul National University, Seoul 08826, Korea.}

\vspace{0.2cm}

E-mail: {\tt hyunwoo1535@snu.ac.kr}

\end{center}

\vspace{0.1cm}

\begin{abstract}

We consider black brane solutions in the STU model, with three R-charges labeled by the parameters $q_1,q_2,q_3$. For equal R-charges $q_1=q_2=q_3$, the solution becomes the Reissner--Nordstr\"om black brane. For unequal R-charges $q_1>q_2>q_3\geq0$, there are three curvature singular points at $R\equiv r^2=-q_i$. Only $R=-q_3$ is the physical singularity, and the other two lie beyond it. Their optical times $T_i$, the complex times of null rays from the boundary to these singular points and back, form a right triangle in the complex $T$-plane. We probe these singular points using three observables, all built from the retarded correlator of $T_{xy}$ at zero spatial momentum: spectral function, OPE coefficients, and QNMs. The exponentially small correction to the spectral function at large real frequency is dominated by $T_1$ and $T_2$, which have the smallest ${\rm Im}\,T_i$. The OPE convergence radius is determined by $T_1$, which has the smallest $|T_i|$. A line of QNMs lies along the direction of $1/T_2$, since $T_2$ has the smallest phase $\arg T_i$. Furthermore, there is a second line of QNMs along the negative imaginary axis, which carries information about $T_3$ corresponding to the physical singularity. We derive these results from a WKB analysis and check them numerically. The three observables are therefore sensitive to different singular points of the geometry.

\end{abstract}

\end{titlepage}

\renewcommand{\thefootnote}{\arabic{footnote}}

\setcounter{footnote}{0}

\renewcommand{\baselinestretch}{1}

\tableofcontents

\renewcommand{\baselinestretch}{1.2}

\section{Introduction}
\label{sec:intro}

In holography \cite{Maldacena:1997re, Gubser:1998bc, Witten:1998qj}, the region behind the horizon of a black hole, including the singularity, can be described by boundary correlators continued analytically from Euclidean signature \cite{Kraus:2002iv}. In the eternal AdS-Schwarzschild black hole, a null geodesic can leave one boundary, bounce off the singularity, and reach the other boundary. Correspondingly, the analytically continued thermal two-point function $G(t)$ is singular at the complex time of this bouncing geodesic \cite{Fidkowski:2003nf, Festuccia:2005pi, Amado:2008hw, Grozdanov:2026cut}. These bouncing singularities appear at leading order in $1/N$, and they are smoothed out at finite $N$ \cite{Komargodski:2026oob, Batra:2026gbq, Dodelson:2026hjm, Liu:2026luf} or at finite 't~Hooft coupling \cite{Dodelson:2025jff}.

Recently, the bouncing geodesic has been identified in three kinds of boundary data. The first is the OPE of the thermal two-point function. Its stress-tensor sector, which contains the identity, the stress tensor and its products, can be computed from a near-boundary expansion of the bulk wave equation \cite{Fitzpatrick:2019zqz}, and the remaining double-trace part can then be fixed by the KMS condition \cite{Buric:2025anb, Buric:2025fye, Barrat:2025twb}. The stress-tensor sector is singular at the complex time of the bouncing geodesic \cite{Ceplak:2024bja}. The second is the spectral function at large real frequency, which receives exponentially small corrections controlled by the bouncing geodesic \cite{Afkhami-Jeddi:2025wra, Jia:2025jbi, Giombi:2026kdz, Jia:2026ryl}. The third is the QNM spectrum. The QNMs at large overtone number are sensitive to the geometry near the singularity, as shown for Kasner interiors \cite{Hartnoll:2026vhu, Xiao:2026pir}, quantum black holes in three dimensions \cite{Coviello:2026tol, Coviello:2026hdc}, and Einstein--Maxwell--dilaton black holes \cite{Hendi:2010gq, Gouteraux:2011ce, Shi:2026cfq}, and their asymptotic spectrum is related to the bouncing singularities \cite{Liu:2026luf, Festuccia:2008zx, Dodelson:2023vrw, Arnaudo:2026tcy}. Thermal OPE coefficients and QNMs can be related directly through thermal Polyakov blocks \cite{Barrat:2026jfg}. The bouncing singularity can also be recovered from the bulk phase shift \cite{Jia:2026pmv}.

When the singularity is spacelike, the geometry near it often takes the Kasner form \cite{Kasner:1921zz, Belinsky:1970ew}, as in black branes with scalar hair \cite{Frenkel:2020ysx, Gao:2023zbd, Arean:2024pzo}. The Kasner exponents can be read off from the boundary data \cite{Hartnoll:2026vhu, Xiao:2026pir, Frenkel:2020ysx, Ceplak:2026ard, Oh:2026dky}. For the Reissner--Nordstr\"om black brane, the singularity is timelike and lies beyond the inner horizon, and the OPE \cite{Ceplak:2025dds} and the spectral function \cite{Chakravarty:2026fdm} still detect a null geodesic bouncing off it.

In the AdS-Schwarzschild and Reissner--Nordstr\"om black branes, the metric has a single singular point, which is the physical singularity in the spacetime, and all the signatures above come from it. In this paper, we consider black branes whose metric has several singular points where the curvature diverges. They are the black branes of the STU model \cite{Behrndt:1998jd, Cvetic:1999xp}, which carry three R-charges of $\mathcal{N}=4$ SYM, with parameters $q_1\geq q_2\geq q_3\geq0$. The metric has curvature singular points at $R\equiv r^2=-q_i$. For equal R-charges, they merge into one, and the solution is the Reissner--Nordstr\"om black brane. For unequal R-charges, there are three singular points, but the spacetime ends at the first one, $R=-q_3$, which is the physical singularity. The other two lie beyond it and are not part of the spacetime. Nevertheless, the bulk wave equation is regular there, so its solutions can be continued to them. Each singular point has its optical time, the complex time of a null ray from the boundary to the point and back. The three optical times $T_1$, $T_2$ and $T_3$ form a right triangle in the complex $T$-plane, as shown in Fig.~\ref{fig:Time}. This naturally raises the question of which singular point each observable detects.

To answer this question, we use three observables, all built from the retarded correlator of $T_{xy}$ at zero spatial momentum: the spectral function, the OPE coefficients, and the QNMs. Unlike a probe operator with an arbitrarily chosen dimension, the stress tensor is present in any CFT, and the bulk dual of $T_{xy}$ obeys the equation of a minimally coupled massless scalar \cite{Kovtun:2004de, Mas:2006dy, Son:2006em}. We derive the signatures of the singular points from a WKB analysis at large frequency, in which a wave sent in from the boundary is reflected very slightly by each singular point, as in \cite{Afkhami-Jeddi:2025wra, Chakravarty:2026fdm}, and we check them numerically.

We find that the singular points that contribute most are different for each observable. The exponentially small correction to the spectral function at large real frequency falls like $e^{-\omega\,{\rm Im}\,T_i}$, so it is dominated by the points with the smallest ${\rm Im}\,T_i$. These are $T_1$ and $T_2$, which have the same imaginary part, so their two terms have the same size and beat. The radius of convergence of the OPE is the distance from $t=0$ to the nearest singularity of $G_R(t)$, which is $|T_1|$, the smallest $|T_i|$. We confirm this from the OPE coefficients alone, without using the optical times. The QNMs appear where $e^{i\omega T_i}$ first becomes of order one as the phase of $\omega$ is rotated, so the complex QNMs lie along the direction of $1/T_2$, where $T_2$ has the smallest $\arg T_i$. In all three cases, the selected points lie beyond the physical singularity. The physical singularity $T_3$ enters the spectral function only through an additional factor on the $T_2$ reflection term, and it sets neither the OPE radius nor the direction of the complex QNMs. It appears at leading order only in the QNMs on the negative imaginary axis, where the terms of $T_2$ and $T_3$ have the same size and interfere. For $q_3=0$, where the physical singularity is a Kasner singularity, these QNMs come in pairs, and $T_1$ governs their exponentially small splitting. The three observables therefore hear different singular points.

This paper is organized as follows. In Section~\ref{sec:STU}, we review the STU black branes and find the optical times of the three curvature singular points. In Section~\ref{sec:Txy}, we introduce the $T_{xy}$ correlator and the three observables, and derive the reflections from the singular points. We then study the spectral function in Section~\ref{sec:tail}, the OPE in Section~\ref{sec:OPE}, and the QNMs in Section~\ref{sec:QNM}. We conclude with remarks and future directions in Section~\ref{sec:conclusion}. Appendix~\ref{sec:A(w)cal} explains how we compute $\mathcal{A}(\omega)$ and the QNMs, and Appendix~\ref{sec:OPEcal} explains how we compute the OPE coefficients.

\section{STU black branes}
\label{sec:STU}

The STU model is the $U(1)^3$ consistent truncation of five-dimensional $\mathcal{N}=8$ $SO(6)$ gauged supergravity, which comes from the $S^5$ reduction of type IIB supergravity \cite{Cvetic:1999xp}. Its three $U(1)$ charges correspond to the three R-charges $U(1)^3\subset SO(6)_R$ of $\mathcal{N}=4$ SYM.

\subsection{Solution}
\label{sec:STUbb}

The planar black brane with three $U(1)$ charges \cite{Behrndt:1998jd} is usually written as
\begin{align}
    ds^2 = - (H_1 H_2 H_3)^{-2/3} f\, dt^2 + (H_1 H_2 H_3)^{1/3} \left( \frac{dr^2}{f} + r^2 d\vec{x}^{2} \right) ,
\end{align}
with
\begin{align}
    H_i = 1+ \frac{q_i}{r^2} , \qquad f = r^2 H_1H_2H_3 -\frac{M}{r^2} , \qquad q_i \geq 0 .
\end{align}
$M$ is a mass parameter, and $\vec{x}$ denotes the three spatial coordinates of the boundary. Without loss of generality, we take
\begin{align}
    q_1 \geq q_2 \geq q_3 \geq 0.
\end{align}

The solution also has three gauge fields and two scalars, but the correlator we study below sees only the metric. The
metric depends on $r$ only through $R=r^2$. With
\begin{align}
    P(R) = \prod_i (R+q_i) , \qquad D(R) = P(R) - MR ,
\end{align}
we have $H_1H_2H_3=P/R^3$ and $f=D/R^2$, so the metric becomes
\begin{align} \label{eq:STUmetric}
    ds^2 = -\frac{D}{P^{2/3}}\, dt^2 + \frac{P^{1/3}}{4D}\, dR^2 + P^{1/3}\, d\vec{x}^{2} .
\end{align}
From now on, we use $R$ instead of $r$. For $q_3>0$, the metric is regular at $R=0$, so the spacetime continues to $R<0$, where $r$ is imaginary, and ends at the singularity $R=-q_3$. The boundary is at $R=\infty$. We use the scaling symmetry of the planar solution to place the outer horizon at $R=1$, so that $D(1)=0$ gives $M=P(1)$. The inverse temperature is given by
\begin{align}
    \beta = \frac{2\pi\sqrt{P(1)}}{D'(1)} .
\end{align}
The $q_i$ are parameters of the metric, and the physical R-charge densities are proportional to $\sqrt{M q_i}$. Equal or unequal $q_i$ therefore mean equal or unequal R-charges.

For $q_1=q_2=q_3=q$, in terms of $\tilde r^{2}=R+q$, we have
\begin{align}
    P = \tilde{r}^6, \qquad D = \tilde{r}^6 -M (\tilde{r}^2 -q  ).
\end{align}
Therefore, \eqref{eq:STUmetric} becomes the
Reissner--Nordstr\"om black brane,
\begin{align}
    ds^2 = -\tilde f\, dt^2 + \frac{d\tilde r^{2}}{\tilde f} + \tilde r^{2} d\vec{x}^{2} , \qquad
    \tilde f = \tilde r^{2} - \frac{M}{\tilde r^{2}} + \frac{Mq}{\tilde r^{4}} ,
\end{align}
with the timelike singularity at $\tilde r=0$ ($R=-q$).

Local thermodynamic stability of the STU black brane \cite{Cvetic:1999rb, Gladden:2024ssb, Gladden:2025glw} holds if and only if
\begin{align}
    2-q_1-q_2-q_3+q_1q_2q_3>0 , \qquad q_1+q_2+q_3<3 .
\end{align}
For a single charge $q_1$ with $q_2=q_3=0$, the stability condition gives $q_1<2$ \cite{Gubser:1998jb, Cvetic:1999ne}. For three equal charges $q_1=q_2=q_3=q$, it gives $q<1$. The wave equation and the analysis below apply to any charges, but the interpretation as $\mathcal{N}=4$ SYM in equilibrium applies only to stable states.

The two examples used throughout this paper, $(q_1, q_2, q_3) = \left( 3,1,\tfrac15 \right)$ and $\left( 2,\tfrac12, 0 \right)$, are locally unstable, and in such states R-charge diffusion modes grow at nonzero momentum \cite{Gladden:2024ssb, Gladden:2025glw}. Nevertheless, we use them because their singular points are well separated, and we find no growing modes in the $T_{xy}$ channel at zero momentum. For the stable states $(1,0,0)$ and $(1.6,0.3,0.02)$, we checked that the OPE and the QNMs follow the same rules, except that $(1,0,0)$ has no QNM line on the imaginary axis, since $T_2=T_3$ there.

\subsection{Three singular points}
\label{sec:3singular}

The zeros of $P$ are singular points, and the zeros of $D$ are horizons. We explain their configuration.

The curvature diverges at each zero of $P$, that is at $R=-q_i$. For example, near a simple zero with $q_i>0$, the Ricci scalar behaves as $(R+q_i)^{-7/3}$. For $q_1>q_2>q_3$, the three zeros are distinct. Coming from the boundary, the spacetime ends at the first one, $R=-q_3$, so this is the physical singularity. The other two, $R=-q_2$ and $R=-q_1$, lie beyond it and are not part of the spacetime. Nevertheless, they matter. They are singular points of the metric, but the wave equation of the next section is regular there, so its solutions continue past $R=-q_3$. We will see that the observables are controlled by different singular points.

Since $D(1)=0$, $D'(1) >0$ from $\beta >0$, and $D(0)=  q_1 q_2 q_3 \geq 0 $, $D(R)$ has a zero in $0\leq R<1$. We call this $R_-$. Also, since $D(-q_1)=Mq_1>0$ and $D(-\infty) =-\infty$, $D(R)$ has a zero in $R<-q_1$, which we call $R_3$. We can write
\begin{align}
    D=(R-1)(R-R_-)(R-R_3) , \qquad 0\leq R_-<1 , \qquad R_3<-q_1 .
\end{align}
$R=1$ is the outer horizon, and for $q_3>0$, $R=R_->0$ is the inner horizon. $R_3$ lies beyond all the singular points, so it is not part of the spacetime.

When $q_3>0$, $D(-q_3)=Mq_3>0$ and $P(R)>0$ for $R>-q_3$. So, $R$ is spacelike near the physical singularity and the physical singularity is timelike, as in the Reissner--Nordstr\"om case. In terms of $r$, the point $R=-q_3$ is the pair $r=\pm i\sqrt{q_3}$. On the other hand, when $q_3=0$, the pair becomes degenerate at $r=0$, and $R_-=0$ as well, so the inner horizon disappears. Then $D<0$ for $0<R<1$, and the singularity at $R=0$ is a spacelike Kasner singularity.

\subsection{Optical times}
\label{sec:optical}

\begin{figure}[H]
    \centering
    \includegraphics[width=\textwidth]{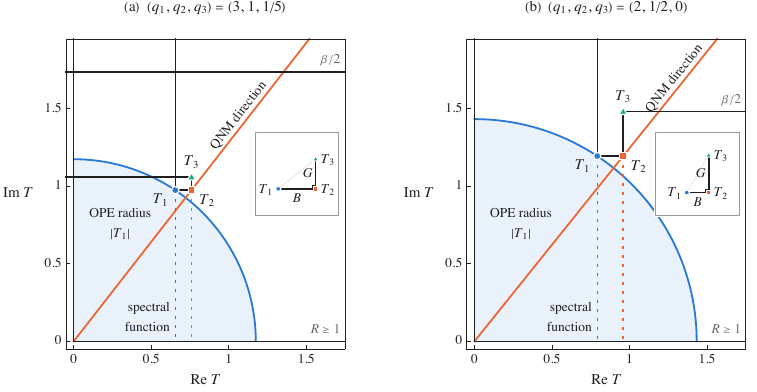}
    \caption{Optical times in the complex $T$-plane. The black lines are the image of the real $R$ axis. The OPE converges inside the disk of radius $|T_1|$. The red line is the direction of $T_2$, and a line of QNMs lies along $1/T_2$. The dotted lines are the Stokes lines of $T_1$ and $T_2$ when $\omega$ is real (Sec.~\ref{sec:reflect}), which control the spectral function at large $\omega$. The insets zoom in on the right triangle formed by $T_1$, $T_2$ and $T_3$.}
    \label{fig:Time}
\end{figure}

From the metric, a radial null ray satisfies $dt = \pm \frac{\sqrt P}{2D} \, dR$. The time it takes to go from the boundary to the singular point $R=-q_i$ and back, is given by
\begin{align} \label{eq:T_c}
    T_i \equiv 2\int_{-q_i}^\infty \frac{\sqrt{P}}{2D}\,dR .
\end{align}
We call this the optical time of $R=-q_i$. The integrand has poles at the horizons where $D=0$, and we pass them through the lower half $R$-plane. This makes $T_i$ complex with ${\rm Im}\,T_i>0$. Replacing $-q_i$ by $R$ in this integral gives a map $R\mapsto T(R)$, and the image of the real $R$ axis is shown by the black lines in Fig.~\ref{fig:Time}.

The exterior $R \geq 1$ maps to the real $T$ axis, and $T\to+\infty$ as $R\to1$. Each pole of the integrand contributes half of its residue. We define
\begin{align}
    \beta_- = \frac{2\pi\sqrt{P(R_-)}}{|D'(R_-)|} , \qquad \beta_3 = \frac{2\pi\sqrt{|P(R_3)|}}{D'(R_3)} ,
\end{align}
in analogy with $\beta$ at the outer horizon $R=1$. We take the absolute value in $\beta_-$ because $D'(R_-)<0$, so that $\beta_-\geq0$, just like $\beta$ and $\beta_3$. Passing $R=1$ shifts $T$ by $i\beta/2$, so $R_-<R<1$ maps to the line ${\rm Im}\,T=\beta/2$. Passing $R=R_-$ shifts $T$ by $-i\beta_-/2$, so $-q_3<R<R_-$ maps to the line ${\rm Im}\,T=\frac12(\beta-\beta_-)$, which ends at $T_3$.

When $-q_2<R<-q_3$, the integrand $\sqrt{P}/2D$ is imaginary, so $T_2$ and $T_3$ are joined by a vertical edge of length $G$. When $-q_1<R<-q_2$, the integrand is real, so $T_1$ and $T_2$ are joined by a horizontal edge of length $B$. When $R<-q_1$, the integrand is imaginary again, so the black line continues vertically. Passing $R=R_3$ shifts $T$ by $-\beta_3/2$. The black line then returns to $T=0$ at $R=-\infty$ along the imaginary axis, since the integrand falls as $|R|^{-3/2}$ and has no poles in the lower half $R$-plane. This gives ${\rm Re}\,T_1=\frac12\beta_3$. Therefore,
\begin{align} \label{eq:triangle}
    T_1=\frac{\beta_3}{2}+\frac{i}{2}\big(\beta-\beta_-\big)-iG , \qquad T_2=T_1+B , \qquad T_3=T_2+iG ,
\end{align}
and the three optical times form a right triangle. For a special case, if $q_3=0$, we have $\beta_-=0$ and hence ${\rm Im}\,T_3=\beta/2$.

For $(q_1,q_2,q_3) = \left( 3,1,\tfrac15 \right)$, we obtain
\begin{align}
    T_1 &= 0.6550170488 + 0.9748769088\, i , \qquad |T_1| = 1.1744923676 , \\
    T_2 &= 0.7597022366 + 0.9748769088\, i , \qquad |T_2| = 1.2359338476 ,  \nn \\
    T_3 &= 0.7597022366 + 1.0585677062\, i , \qquad |T_3| = 1.3029631909 ,  \nn
\end{align}
with $B = 0.1046851878$ and $G = 0.0836907974$.

For $(q_1,q_2,q_3) = \left( 2,\tfrac12,0 \right)$, we obtain
\begin{align}
    T_1 &= 0.7916069411 + 1.1947240304\, i , \qquad |T_1| = 1.4331807486 , \\
    T_2 &= 0.9588164034 + 1.1947240304\, i , \qquad |T_2| = 1.5318924258 ,  \nn \\
    T_3 &= 0.9588164034 + 1.4809609794\, i , \qquad |T_3| = 1.7642489388 ,  \nn
\end{align}
with $B = 0.1672094623$ and $G = 0.2862369490$.

\section{The $T_{xy}$ correlator}
\label{sec:Txy}

We study the retarded correlator of the transverse stress-tensor component $T_{xy}$ at zero spatial momentum, where $x$ and $y$ are two components of $\vec{x}$ in \eqref{eq:STUmetric}. In any four-dimensional CFT, the stress tensor is a universal operator of dimension $ \Delta=4$. Thus, unlike a probe operator with an arbitrarily chosen mass or dimension, $T_{xy}$ is an operator that is intrinsically present in $ \mathcal N=4$ SYM.

This channel is particularly simple in the charged STU background. At zero spatial momentum, its bulk dual, the tensor metric fluctuation ${h^x}_y$, decouples from the gauge fields and scalars and obeys the equation of a minimally coupled massless scalar \cite{Kovtun:2004de, Mas:2006dy, Son:2006em}. Other channels generally involve coupled fluctuations at finite density. The \(T_{xy}\) channel has been extensively studied through shear viscosity, quasinormal modes, and spectral functions \cite{Kovtun:2004de, Policastro:2001yc, Starinets:2002br, Nunez:2003eq, Kovtun:2005ev, Teaney:2006nc, Kovtun:2006pf}, and more recently in studies of boundary signatures of black-hole singularities \cite{Afkhami-Jeddi:2025wra, Jia:2025jbi, Chakravarty:2026fdm}. Here, the $T_{xy}$ channel provides a clean setting in which to compare how the different singular points of the STU geometry appear in the spectral function, the OPE, and the QNM spectrum.

\subsection{Wave equation}

At zero spatial momentum, the transverse graviton ${h^x}_y=\phi(R)\,e^{-i\omega t}$ has spin 2 under rotations of $\vec{x}$, so it decouples from the gauge fields and the scalars, and it is a minimally coupled massless scalar, so its wave equation is given by
\begin{align}
    \partial_\mu \big( \sqrt{-g}\, g^{\mu\nu} \partial_\nu {h^x}_y \big) = 0 .
\end{align}
Since the metric is diagonal and depends only on $R$, and ${h^x}_y$ does not depend on $\vec{x}$, this becomes
\begin{align}
    \partial_R \big( \sqrt{-g}\, g^{RR} \partial_R \phi  \big) + \omega^2 \sqrt{-g}\, (-g^{tt})\, \phi = 0 .
\end{align}
From \eqref{eq:STUmetric}, we have
\begin{align}
    -g = \frac{D}{P^{2/3}}\cdot \frac{P^{1/3}}{4D} \cdot (P^{1/3})^3 = \frac{P^{2/3}}{4},
\end{align}
since $\vec{x}$ is three dimensional, and
\begin{align}
    \sqrt{-g}\,g^{RR}=2D , \qquad \sqrt{-g}\,(-g^{tt})=\frac{P}{2D} .
\end{align}
So, the wave equation becomes
\begin{align} \label{eq:waveR}
    \partial_R (D \, \partial_R  \phi)=-\frac{\omega^2P}{4D}\,\phi .
\end{align}
We introduce the coordinate $x$ by
\begin{align}
    dx=-\frac{dR}{2D} , \qquad x= -  \int_\infty^R \frac{dR'}{2D(R')},
\end{align}
which vanishes at the boundary, increases inwards and diverges logarithmically at the horizon. In terms of $x$, the wave equation is written in the Schr\"{o}dinger form,
\begin{align} \label{eq:wavex}
    \phi''  +\omega^2P\,\phi=0,
\end{align}
where $'=\frac{d}{dx}$.

\subsection{Observables}
\label{sec:obs}

Near the boundary, the two solutions of the wave equation behave as $\phi_A\simeq1$ and $\phi_B\simeq R^{-2}$. They are the non-normalizable and the normalizable mode. We write the solution that is infalling at the horizon as
\begin{align}
    \phi_{\rm in}=\mathcal{A}(\omega)\,\phi_A+\mathcal{B}(\omega)\,\phi_B .
\end{align}

Since $\mathcal{A}$ is the source and $\mathcal{B}$ is the response, up to normalization and contact terms, the retarded correlator of $T_{xy}$ at zero spatial momentum,
\begin{align}
    G_R(t)=i\theta(t)\,\big\langle[T_{xy}(t),T_{xy}(0)]\big\rangle_\beta , \qquad G_R(\omega)=\int dt\,e^{i\omega t}\,G_R(t) ,
\end{align}
is given by \cite{Son:2002sd}
\begin{align}
    G_R(\omega)=\frac{\mathcal{B}(\omega)}{\mathcal{A}(\omega)} .
\end{align}
We study three observables built from it. The first is the spectral function $\rho(\omega)={\rm Im}\,G_R(\omega)$ for real $\omega$. In Appendix~\ref{sec:A(w)cal}, when $\phi_{\rm in}$ is normalized as $|\phi_{\rm in}|\to1$ at the horizon, we show that
\begin{align} \label{eq:rhoA}
    \rho = \frac{\omega\sqrt{P(1)}}{4|\mathcal{A}(\omega) |^2} ,
\end{align}
and that the vacuum value is $\rho_{\rm vac}=\pi\omega^4/32$. At large real $\omega$, we study the difference $\Delta\rho=\rho-\rho_{\rm vac}$. The second is the set of OPE coefficients, the coefficients of the expansion of $G_R(t)$ near $t=0$. The third is the set of QNMs, the poles of $G_R(\omega)$. They are the solutions of $\mathcal{A}(\omega)=0$.

\subsection{Reflection from the singular points}
\label{sec:reflect}

We now find $\mathcal{A}$ and $\mathcal{B}$ at large $\omega$. The idea is that a wave sent in from the boundary is reflected very slightly by the singular points, and the reflected wave modifies $\mathcal{A}$ and $\mathcal{B}$. The same idea was used for the AdS-Schwarzschild black brane in \cite{Afkhami-Jeddi:2025wra} and for the Reissner--Nordstr\"om black brane in \cite{Chakravarty:2026fdm}, each with a single singular point.

At large $\omega$ we solve \eqref{eq:wavex} by the WKB method. Writing $\phi=e^{\Phi}$, the equation becomes
$\Phi'' + \Phi'^2+\omega^2P=0$. We expand
\begin{align}
    \Phi' = i\omega\,y_0 + y_1 + O(1/\omega)
\end{align}
and collect powers of $\omega$:
\begin{align}
    y_0=\pm\sqrt{P} , \qquad y_1=-\frac{P'}{4P} .
\end{align}
The two WKB solutions are therefore
\begin{align} \label{eq:fpm}
    f_\pm = P^{-1/4}\, e^{\pm i\omega S} , \qquad S(x)=\int_0^x\sqrt{P}\,dx' .
\end{align}
Since $dS = \sqrt{P} \, dx = -\frac{\sqrt{P}}{2D} \, dR$, $S$ is the tortoise coordinate measured from the boundary. It vanishes at the boundary and grows inward. With the time dependence $e^{-i\omega t}$, $f_+$ moves inward, so near the horizon
\begin{align}
    \phi_{\rm in}\propto f_+ .
\end{align}
The WKB approximation fails near the zeros of $P$ and near the boundary. On the real section $R\geq1$, between the horizon and the boundary, we have $P>0$, so WKB fails there only near the boundary, which we treat separately below. The WKB series of $f_+$ never produces $f_-$: to any order in $1/\omega$, an infalling wave stays infalling. Any reflection is therefore exponentially small, and it must come from the zeros of $P$, where WKB fails.

The zeros of $P$ are not on the real section $R\geq1$, but those with $q_c>0$ are regular points of the wave equation, so we can analytically continue $\phi$ to them through the complex plane. In the complex $S$ plane, the singular point $R=-q_c$ sits at
\begin{align}
    S_c = \int_0^{x_c} \sqrt{P}\,dx = \int_{-q_c}^\infty  \frac{\sqrt{P}}{2D}dR   = \frac{T_c}{2} , \qquad {\rm Im}\,S_c > 0 ,
\end{align}
where $T_c$ is its optical time \eqref{eq:T_c}. The physical path, the real $S$ axis, passes below $S_c$ since ${\rm  Im} \, S_c>0$. We show next that near $S_c$ the infalling wave acquires a reflected component of size $e^{2i\omega S_c}=e^{i\omega T_c}$.

At $x=x_c$, we have $P=0$. Near $x_c$, we write $P\simeq CX^\ell$ with $X=x-x_c$. The local equation $\partial_X^2 \phi + \omega^2CX^\ell\phi=0$ is solved exactly by
\begin{align} \label{eq:localBessel}
    \phi=\sqrt{X}\,Z_\nu(\xi) , \qquad \nu=\frac{1}{\ell+2} , \qquad \xi=\frac{2\omega\sqrt{C}}{\ell+2}\,X^{\frac{\ell+2}{2}} ,
\end{align}
where $Z_\nu$ is any Bessel function of order $\nu$. For a simple zero with $q_c>0$, we have $D(R=-q_c)=Mq_c\neq0$, so $dx/dR=-1/(2D)$ is finite and nonzero there. Hence $X\simeq-\frac{R+q_c}{2Mq_c}$ near $x_c$, so $\ell=1$ and $\nu=\frac13$. In this case the solutions \eqref{eq:localBessel} are Airy functions.

Near $x_c$, $dS = \sqrt{P} \, dx \simeq \sqrt{C} \, X^{\ell/2}dX$, so $\xi\simeq\omega(S-S_c)$. For $|\xi| \gg 1$,
\begin{align} \label{eq:Hankelnu}
    H^{(1,2)}_\nu(\xi)\simeq\sqrt{\frac{2}{\pi\xi}}\,e^{\pm i\left(\xi-\frac{\nu\pi}{2}-\frac{\pi}{4}\right)} ,
\end{align}
and since $\sqrt{X}\,\xi^{-1/2} \propto  X^{-\frac{\ell}{4}} \propto P^{-1/4}$, the solutions \eqref{eq:localBessel} reduce to the WKB waves:
\begin{align}
    \sqrt{X} \, H^{(1)}_\nu(\xi) \propto f_+ , \qquad \sqrt{X} \, H^{(2)}_\nu(\xi)\propto f_- .
\end{align}
The WKB waves hold far from the singular point and the local solutions hold near it. At large $\omega$ these two regions overlap, and the two descriptions are matched there.

\begin{figure}[H]
    \centering
    \captionsetup[subfigure]{singlelinecheck=false, justification=raggedright, skip=2pt, margin={0.212\linewidth,0pt}}
    \begin{subfigure}{0.49\textwidth}
        \caption{$q_3>0$}
        \includegraphics[width=\textwidth]{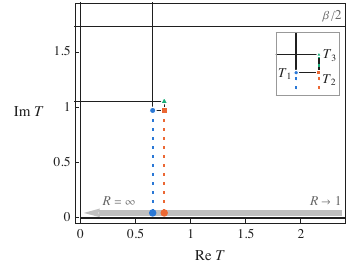}
    \end{subfigure}
    \hfill
    \begin{subfigure}{0.49\textwidth}
        \caption{$q_3>0$}
        \includegraphics[width=\textwidth]{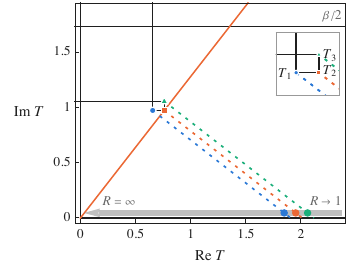}
    \end{subfigure}
    \begin{subfigure}{0.49\textwidth}
        \caption{$q_3 >0$}
        \includegraphics[width=\textwidth]{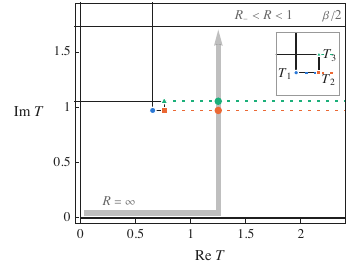}
    \end{subfigure}
    \hfill
    \begin{subfigure}{0.49\textwidth}
        \caption{$q_3=0$}
        \includegraphics[width=\textwidth]{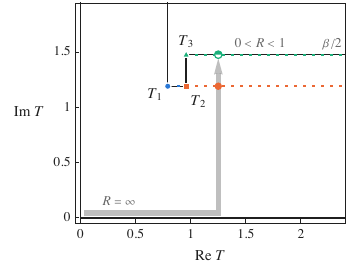}
    \end{subfigure}
    \caption{Stokes lines (dotted) of $T_1$, $T_2$ and $T_3$ for $(3,1,\tfrac15)$, for (a) real $\omega$, (b) $\arg\omega=-\arg T_2$ and (c) $\omega=-iy$, and (d) for $(2,\tfrac12,0)$ and $\omega=-iy$. The black lines are those of Fig.~\ref{fig:Time}, and the insets in (a)--(c) zoom in on $T_1$, $T_2$ and $T_3$. The gray arrow is the path along which $\phi_{\rm in}$ in (a) and (b), or $\phi_B$ in (c) and (d), is followed from where it is fixed, and a reflected wave is added at each dot, where the path crosses a Stokes line. In (a), the Stokes line of $T_3$ runs along the vertical edge into $T_2$. In (b), the red line is where $|e^{i\omega T}|=1$, and it passes through $T_2$. In (c) and (d), the Stokes line of $T_1$ runs along the horizontal edge into $T_2$. In (d), $T_3$ lies on the line ${\rm Im}\,T=\beta/2$, so the path ends on the Stokes line of $T_3$ (half-filled dot) instead of crossing it.}
    \label{fig:Stokes}
\end{figure}

On the horizon side of $S_c$, $\phi_{\rm in}$ is a pure $f_+$ up to exponentially small terms. Near $x_c$ it is
therefore $\sqrt{X}\,H^{(1)}_\nu(\xi)$ with $\arg\xi\approx0$. Following the real axis toward the boundary, the path
passes below $S_c$, and $\arg(S-S_c)$ changes from $0$ to $-\pi$. On the boundary side we therefore write
\begin{align}
    \xi\simeq\xi_0 e^{-i\pi} , \qquad \xi_0=\omega(S_c-S) ,
\end{align}
with $\xi_0$ approximately real and positive, and use the continuation formula \cite[\S10.11]{Olver:2010ouy}
\begin{align} \label{eq:Hcont}
    H^{(1)}_\nu(\xi_0 e^{-i\pi}) &= 2\cos(\nu\pi) \, H^{(1)}_\nu (\xi_0) + e^{-i\nu\pi} \, H^{(2)}_\nu (\xi_0) \\
    &= \sqrt{\frac{2}{\pi \xi_0}} \left[ 2 \cos (\nu \pi) \, e^{ -i \left( \frac{\nu \pi}{2} + \frac{\pi}{4} \right) } e^{i\xi_0 } + e^{ -i\nu \pi
    } \, e^{ i \left( \frac{\nu \pi}{2} + \frac{\pi}{4} \right) } e^{-i\xi_0 }    \right] . \nn
\end{align}
Since $\xi_0=\omega(S_c-S)$,
\begin{align}
    e^{i\xi_0} = e^{i\omega S_c} \, e^{-i\omega S} =  e^{i\omega S_c}P^{1/4} f_- , \qquad
    e^{-i\xi_0} = e^{-i\omega S_c} \, e^{+i\omega S} =  e^{-i\omega S_c} P^{1/4} f_+ .
\end{align}
The ratio of the two coefficients of $f_-, \, f_+$ on the boundary side is
\begin{align}
    \frac{(f_-\ \text{coefficient})}{(f_+\ \text{coefficient})}
    &= \frac{2\cos(\nu\pi)\,e^{-i\left(\frac{\nu\pi}{2}+\frac{\pi}{4}\right)}\,e^{i\omega S_c}}
           {e^{-i\nu\pi}\,e^{i\left(\frac{\nu\pi}{2}+\frac{\pi}{4}\right)}\,e^{-i\omega S_c}}  \\
    &= -2i\cos(\pi\nu)\,e^{i\omega T_c} .  \nn
\end{align}
On the boundary side of $S_c$, the infalling solution is therefore
\begin{align} \label{eq:reflect}
    \phi_{\rm in}\propto f_+  - 2i \cos(\pi \nu_c) \, e^{i\omega T_c}\,f_-  ,
\end{align}
where $\nu_c$ is the value of $\nu$ at $R=-q_c$. The reflected wave is switched on where the path crosses the vertical line ${\rm Re}\,S={\rm Re}\,S_c$ below $S_c$. On this line $e^{2i\omega(S-S_c)}$ is real, so one WKB wave is maximally dominant over the other. This line is the Stokes line of the singular point. In terms of $T=2S$, it is the vertical line ${\rm Re}\,T={\rm Re}\,T_c$ below $T_c$. Fig.~\ref{fig:Stokes}(a) shows the Stokes lines of $T_1$, $T_2$ and $T_3$ together with the path. The reflected wave is switched on only where the path crosses a Stokes line. So the result depends only on which Stokes lines the path crosses, and in which direction it crosses each of them. It does not depend on the shape of the path, as long as the path does not cross the black lines of Fig.~\ref{fig:Time}.

Each singular point produces its own reflection. Every reflection is exponentially small, so terms with two reflections are of order $e^{-\omega({\rm Im}\,T_c+{\rm Im}\,T_{c'})}$ and can be dropped:
\begin{align} \label{eq:reflectsum}
    \phi_{\rm in} \propto f_+ - \Big(\sum_c 2i \cos (\pi \nu_c) \,e^{i\omega T_c}\Big) f_- .
\end{align}

The local computation above assumes that the Stokes line reaches the real axis without meeting another singular point. We call a singular point visible if the vertical segment from $T_c$ down to the real $T$ axis does not meet another black line of Fig.~\ref{fig:Time}. Only visible points contribute their own term to \eqref{eq:reflectsum}. When $\omega$ is real, $T_1$ and $T_2$ are visible, since the black lines other than the real $T$ axis lie at ${\rm Im}\,T\geq{\rm Im}\,T_2$ or on ${\rm Re}\,T=0$. Since $T_2-T_1=B$ is real, they contribute at the same order $e^{-\omega\,{\rm Im}\,T_1}$. $T_3$ is not visible, since the segment below $T_3$ is the vertical edge of length $G$ itself, which ends at $T_2$. We say that $T_3$ is shadowed. Its Stokes line ends at $T_2$ instead of reaching the real $T$ axis, so the local computation above does not apply to it. In Sec.~\ref{sec:tail}, we treat $T_2$ and $T_3$ together, and find that $T_3$ still has its own reflection term, with half of its amplitude.

It remains to match to the boundary. Near the boundary,
\begin{align}
    D(R) \simeq R^3 , \qquad  x = -\int_\infty^R \frac{dR'}{2D(R')} \simeq \frac{1}{4R^2} , \qquad R \simeq (4x)^{-1/2},
\end{align}
so $P \simeq R^3 \simeq (4x)^{-3/2}$. With $z=1/\sqrt{R}$, we have
\begin{align}
    S\simeq(4x)^{1/4}=z , \qquad P^{-1/4}\simeq z^{3/2} .
\end{align}
For $\psi=P^{1/4}\phi$, \eqref{eq:waveR} then becomes
\begin{align}
    \psi_{zz}+\Big(\omega^2-\frac{15}{4z^2}\Big)\psi=0 .
\end{align}
The WKB solutions $f_\pm$ are $\psi=e^{\pm i\omega z}$, which neglect the $1/z^2$ term. So WKB fails for $\omega z\lesssim1$, but the solutions are the vacuum ones,
$z^2H^{(1,2)}_2(\omega z)=R^{-1}H^{(1,2)}_2(\omega/\sqrt{R})$. WKB holds for $\omega z\gg1$ and the vacuum form holds for $\omega z^3\ll1$, since $S=z+O(z^3)$, so at large $\omega$ the two can be matched for $1/\omega\ll z\ll\omega^{-1/3}$. Using
\begin{align}
    z^2 H^{(1,2)}_2(\omega z) \simeq \sqrt{\frac{2}{\pi\omega}}\; z^{3/2}\, e^{\pm i\left(\omega z-\frac{5\pi}{4}\right)} ,
\end{align}
we have
\begin{align}
    f_+ = \sqrt{\frac{\pi\omega}{2}}\,e^{\frac{5\pi i}{4}}\,z^2H^{(1)}_2(\omega z) , \qquad
    f_- = \sqrt{\frac{\pi\omega}{2}}\,e^{-\frac{5\pi i}{4}}\,z^2H^{(2)}_2(\omega z) ,
\end{align}
and \eqref{eq:reflectsum} becomes
\begin{align} \label{eq:rhotilde}
    \phi_{\rm in}\propto z^2\Big[H^{(1)}_2(\omega z)+\tilde\rho\,H^{(2)}_2(\omega z)\Big] , \qquad
    \tilde\rho =  -2\sum_c\cos(\pi\nu_c)\,e^{i\omega T_c} .
\end{align}

For $1/\omega\ll z\ll\omega^{-1/3}$, $\phi_A$ and $\phi_B$ also take the vacuum form of Appendix~\ref{sec:A(w)cal}, $z^2J_2(\omega z)=\frac{\omega^2}{8}\phi_B$ and
$z^2Y_2(\omega z)=-\frac{4}{\pi\omega^2}\phi_A$. Hence
\begin{align} \label{1pmtilderho}
    z^2H^{(1,2)}_2(\omega z) = \frac{\omega^2}{8}\,\phi_B \mp \frac{4i}{\pi\omega^2}\,\phi_A ,
    \qquad
    \mathcal{A}\propto 1-\tilde\rho , \qquad \mathcal{B}\propto 1+\tilde\rho ,
\end{align}
and
\begin{align} \label{eq:BAtail}
    \frac{\mathcal{B}}{\mathcal{A}} = \frac{i\pi\omega^4}{32}\,\frac{1+\tilde\rho}{1-\tilde\rho} .
\end{align}

\section{Spectral function}
\label{sec:tail}

From \eqref{eq:BAtail}, we have
\begin{align}
    \frac{\rho}{\rho_{\rm vac}} =\frac{ {\rm Im} (\mathcal{B}/\mathcal{A} ) }{\pi \omega^4 /32}  = {\rm Re}\,\frac{1+\tilde\rho}{1-\tilde\rho} .
\end{align}
Since $\tilde\rho$ of \eqref{eq:rhotilde} is exponentially small,
\begin{align}
    \frac{\Delta \rho}{\rho_{\rm vac}} = \frac{\rho-\rho_{\rm vac}}{ \rho_{\rm vac} } = {\rm Re} \frac{2\tilde \rho}{1-\tilde \rho} \simeq 2  \, {\rm Re} \, \tilde\rho ,
\end{align}
and inserting $\tilde \rho$ gives
\begin{align} \label{eq:tailLO}
    \frac{\Delta\rho}{\rho_{\rm vac}} \simeq {\rm Re}\sum_c A_c\, e^{i\omega T_c} , \qquad A_c=-4\cos(\pi\nu_c) ,
\end{align}
where the sum runs over the visible points $T_1$ and $T_2$. We call this exponentially small correction of $\rho$ to $\rho_{\rm vac}$ the spectral tail. The points $T_1$ and $T_2$ have the smallest ${\rm Im}\,T_c$, so the tail falls like $e^{-\omega\,{\rm Im}\,T_1}$.

The amplitude $A_c$ depends only on the order $\ell$ of the zero of $P$, through $\nu_c=1/(\ell+2)$. A simple zero has $\nu=\frac13$ and $A=-2$, a double zero has $\nu=\frac14$ and $A=-2\sqrt2$, and a triple zero has $\nu=\frac15$ and $A=-4\cos(\pi/5)$. Since $R=-q_1$ and $R=-q_2$ are simple zeros, $A_1=A_2=-2$. For the Reissner--Nordstr\"om black brane, the three zeros merge into a triple zero, and its amplitude agrees with \cite{Chakravarty:2026fdm}.

Since $T_2-T_1=B$ is real, the two terms from $T_1, T_2$ have the same size and they beat as
\begin{align} \label{eq:beat}
    \frac{\Delta\rho}{\rho_{\rm vac}} &\simeq {\rm Re}\left[-2e^{i\omega T_1}-2e^{i\omega T_2}\right]
    \\
    &= -4\,e^{-\omega\,{\rm Im}\,T_1}\cos\!\Big(\omega\,\frac{{\rm Re}\,T_1+{\rm Re}\,T_2}{2}\Big)\cos\!\Big(\frac{\omega B}{2}\Big)   \nn .
\end{align}
The first cosine is the fast oscillation, with period $4\pi/({\rm Re}\,T_1+{\rm Re}\,T_2)$, which is $8.9$ for $(3,1,\tfrac15)$ and $7.2$ for $(2,\tfrac12,0)$. Since $B\ll{\rm Re}\,T_1+{\rm Re}\,T_2$, the second cosine varies slowly and gives its envelope, with period $2\pi/B$, which is $60.0$ for $(3,1,\tfrac15)$ and $37.6$ for $(2,\tfrac12,0)$.

The shadowed point $T_3$ enters through the $T_2$ term, since ${\rm Re}\, T_2 = {\rm Re} \, T_3$. In Sec.~\ref{sec:reflect}, we treated each zero of $P$ separately. In the $T$-plane, the local solutions around $T_2$ are needed in a region of size of order $1/\omega$, and $T_3$ is at distance $G$ from $T_2$, so this requires $\omega G\gg1$. For $(2,\tfrac12,0)$, we have $G=0.286$, so in the range $16\leq\omega\leq120$ of our numerics $\omega G$ runs from $4.6$ to $34$. Then $T_3$ lies well outside the region of size $1/\omega$ around $T_2$, so we use \eqref{eq:beat}. For $(3,1,\tfrac15)$, we have $G=0.0837$, so in this range $\omega G$ only runs from $1.3$ to $10$, and we have to treat the two zeros together.

Near them, $P$ is quadratic in $x$. With $X=x-\frac12(x_2+x_3)$ and $\delta=\frac12|x_2-x_3|$, we write $P\simeq C(X^2-\delta^2)$. Integrating $\sqrt{P}$ from $x_2$ to $x_3$ gives $S_3-S_2=iG/2$, so $G=\pi\sqrt{C}\,\delta^2$. In terms of $\zeta=(4\omega^2C)^{1/4}X$, \eqref{eq:wavex} becomes
\begin{align}
    \partial_\zeta^2\phi+\Big(\frac{\zeta^2}{4}-a\Big)\phi=0 , \qquad a=\frac{\omega G}{2\pi} .
\end{align}
This local equation is solved exactly by parabolic cylinder functions \cite[\S12.14]{Olver:2010ouy}, which replace the Bessel functions of \eqref{eq:localBessel}. Repeating the calculation of Sec.~\ref{sec:reflect}, the reflected wave of $T_2$ in \eqref{eq:reflectsum} is multiplied by a factor $W(\omega G)$, and we obtain
\begin{align} \label{eq:tailW}
    \frac{\Delta\rho}{\rho_{\rm vac}} \simeq {\rm Re}\left[-2\,e^{i\omega T_1}-2\,W(\omega G)\,e^{i\omega T_2}\right] ,
\end{align}
where
\begin{align} \label{eq:defofW}
    W(g) := \frac{\sqrt{2\pi}\,e^{- g/4} \, (g/2\pi e)^{-ia}}{\Gamma\!\left(\frac12 - \frac{ig}{2\pi}\right)} , \qquad a=\frac{g}{2\pi} .
\end{align}
By Stirling's formula, $W(\omega G)\to1$ for $\omega G\gg1$, so $T_2$ reflects alone, as we explained. For $G\to0$, the two zeros merge into a double zero with $\ell=2$ and $\nu=\frac14$, and $-2W\to-2\sqrt2$ reproduces its amplitude.

More precisely, since $|W(g)|^2=1+e^{-g}$, we have $W(\omega G) \simeq e^{i\theta} \big(1+\tfrac12e^{-\omega G}\big)$ for $\omega G\gg1$, where the phase $\theta$ is real and vanishes as $\omega G\to\infty$. Keeping the leading coefficient of each exponential, \eqref{eq:tailW} becomes
\begin{align}
    \frac{\Delta \rho}{\rho_{\rm vac}} \simeq {\rm Re} \left[ -2e^{i\omega T_1} -2 e^{i \omega T_2} - e^{i\omega T_3}  \right].
\end{align}
The coefficient of $e^{i\omega T_3}$ is half of $A_3=-2$, as for the QNMs in \eqref{eq:doubleE} below. We can interpret this as follows: For real $\omega$, the Stokes line of $T_3$ runs along the vertical edge into $T_2$ (Fig.~\ref{fig:Stokes}(a)). If $\omega$ has a small negative imaginary part, this line passes to the right of $T_2$ and reaches the real $T$ axis, so $T_3$ becomes visible and fully contributes $A_3 e^{i\omega T_3}$. If $\omega$ has a small positive imaginary part, this line passes to the left of $T_2$ and meets the black line from $T_1$ to $T_2$, so $T_3$ does not contribute. Real $\omega$ lies between the two cases, and \eqref{eq:defofW} gives the intermediate value $\frac12A_3$, as expected on a Stokes line \cite{Barry:1989zz}. Therefore, we can think the physical singularity has its own reflection term with half its amplitude.\footnote{I thank Joydeep Chakravarty for pointing this out.}

We now compare these formulas \eqref{eq:beat} and \eqref{eq:tailW} with numerics. From \eqref{eq:rhoA},
\begin{align}
    \frac{\Delta \rho}{\rho_{\rm vac}} = \frac{8\sqrt{P(1)}}{\pi \omega^3 |\mathcal{A}(\omega) |^2} -1 ,
\end{align}
so we only need $\mathcal{A}(\omega)$, which we compute as explained in Appendix~\ref{sec:A(w)cal}. Since $\Delta\rho/\rho_{\rm vac}$ from \eqref{eq:beat} and \eqref{eq:tailW} falls like $e^{-\omega\,{\rm Im}\,T_1}$, we plot $e^{\omega\,{\rm Im}\,T_1}\,\Delta\rho/\rho_{\rm vac}$, which stays of order one.

\begin{figure}[H]
    \centering
    \includegraphics[width=\textwidth]{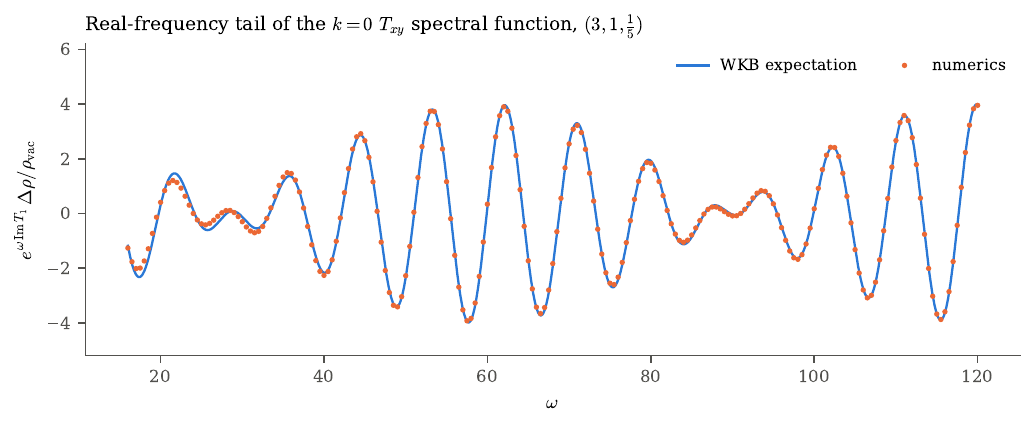}
    \caption{$e^{\omega \, {\rm Im} \, T_1 } \, \Delta \rho / \rho_{\rm vac}$ for $(3,1,\tfrac15)$. The points are the numerics and the line is \eqref{eq:tailW}, with no fitted parameter.}
    \label{fig:tail1}
\end{figure}

\begin{figure}[H]
    \centering
    \includegraphics[width=\textwidth]{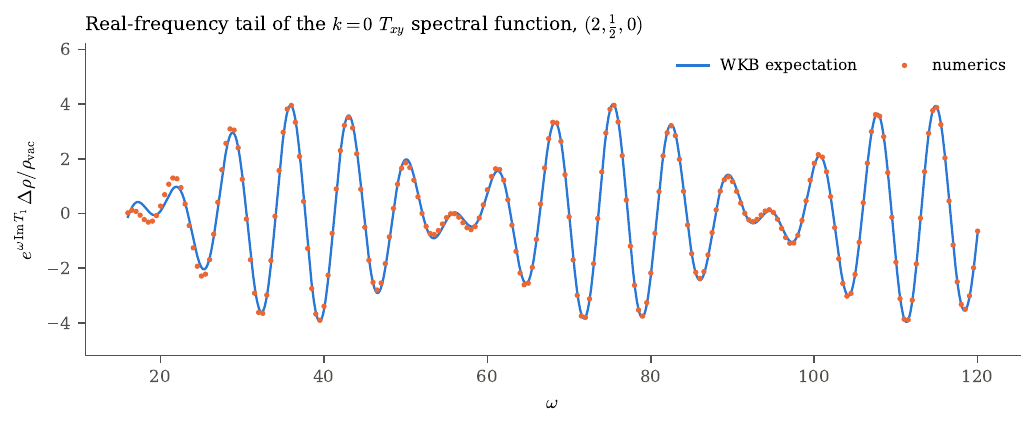}
    \caption{$e^{\omega\,{\rm Im}\,T_1}\,\Delta\rho/\rho_{\rm vac}$ for $(2,\tfrac12,0)$. The points are the numerics and the line is \eqref{eq:beat}, with no fitted parameter.}
    \label{fig:tail2}
\end{figure}

Fig.~\ref{fig:tail1} compares the numerics for $(3,1,\tfrac15)$ with \eqref{eq:tailW}, and Fig.~\ref{fig:tail2} compares the numerics for $(2,\tfrac12,0)$ with \eqref{eq:beat}. These formulas have no fitted parameter, and the two agree, including the beats. To test the amplitudes, we also tried fitting the data with $\omega\geq55$, leaving $A_1$ and $A_2$ in \eqref{eq:tailLO} free and allowing corrections up to $1/\omega^3$ to each term. For $(3,1,\tfrac15)$, with the factor $W$, we obtain $A_1=-2.0004+0.0005i$ and $A_2=-1.9998-0.0004i$. For $(2,\tfrac12,0)$, we obtain $A_1=-2.0000+0.0001i$ and $A_2=-2.0001-0.0001i$.

Therefore, the spectral function at large real $\omega$ is controlled by $T_1$ and $T_2$.

\section{OPE convergence radius}
\label{sec:OPE}

At short times, the retarded correlator is a vacuum term plus a power series,
\begin{align} \label{eq:OPE}
    G_R(t) = \frac32 \, t^{-5} + \sum_{m\geq1}\gamma_m\,t^{2m-1} ,
\end{align}
as we show in Appendix~\ref{sec:OPEcal}, where we also explain how to compute $\gamma_m$. The radius of convergence of the series is the distance from $t=0$ to the nearest singularity of $G_R(t)-\frac32t^{-5}$. The spectral function tells us about these singularities. Since $G_R(t)$ vanishes for $t<0$ and is real, the spectral function is
\begin{align}
    \rho(\omega)={\rm Im}\int_0^\infty dt\,e^{i\omega t}\,G_R(t)=\int_0^\infty dt\,\sin(\omega t)\,G_R(t) ,
\end{align}
and inverting this sine transform gives
\begin{align}
    G_R(t)=\frac{2}{\pi}\int_0^\infty d\omega\,\sin(\omega t)\,\rho(\omega) , \qquad t>0 .
\end{align}
Using
\begin{align} \label{eq:w4int}
    \int_0^\infty d\omega\,\omega^4\,e^{i\omega\tau}=\frac{24\,i}{\tau^5} , \qquad {\rm Im}\,\tau>0 ,
\end{align}
the vacuum part $\rho_{\rm vac} = \pi\omega^4/32$ gives the vacuum term $\frac32t^{-5}$, so
\begin{align} \label{eq:GRsine}
    G_R(t)-\frac32t^{-5}=\frac{2}{\pi}\int_0^\infty d\omega\,\sin(\omega t)\,\Delta\rho(\omega) , \qquad t>0 .
\end{align}
Since $\Delta\rho$ falls like $e^{-\omega\,{\rm Im}\,T_1}$, this integral converges and is analytic for $|{\rm Im}\,t|<{\rm Im}\,T_1$. Both $T_1$ and $T_2$ lie on the edge of this strip (${\rm Im}\,T_1={\rm Im}\,T_2$), and the two terms of the spectral tail \eqref{eq:tailLO} produce singularities there. Since $\tilde\rho$ in \eqref{eq:rhotilde} has no power of $\omega$ in front of $e^{i\omega T_c}$, these terms keep the $\omega^4$ of $\rho_{\rm vac}$,
\begin{align}
    \Delta\rho\simeq\frac{\pi\omega^4}{32}\,{\rm Re}\sum_{c=1,2}A_c\,e^{i\omega T_c} .
\end{align}
Writing $\sin(\omega t)$ and the real part in exponentials, and using that $A_c$ is real,
\begin{align}
    \sin(\omega t)\,{\rm Re}\big[A_c\,e^{i\omega T_c}\big]=\frac{A_c}{4i}\Big(e^{i\omega(T_c+t)}-e^{i\omega(T_c-t)}+e^{i\omega(t-\bar T_c)}-e^{-i\omega(t+\bar T_c)}\Big) .
\end{align}
Inside the strip, each exponential is $e^{i\omega\tau}$ with ${\rm Im}\,\tau>0$, so \eqref{eq:w4int} applies to each of them, and inserting the term of $T_c$ into \eqref{eq:GRsine} gives
\begin{align} \label{eq:fourpoles}
    C_c\Big[(t-T_c)^{-5}+(t+T_c)^{-5}+(t-\bar T_c)^{-5}+(t+\bar T_c)^{-5}\Big] , \qquad C_c=\frac{3A_c}{8} .
\end{align}
So $T_1$ and $T_2$ are singularities of $G_R(t)$, where the vacuum singularity $\frac32t^{-5}$ reappears with the coefficient $C_c=-\frac34$. The straight lines from $t=0$ to $T_1$ and $T_2$ lie inside the strip, so both are on the principal sheet, even though $R=-q_1$ and $R=-q_2$ lie beyond the physical singularity.

These singularities determine $\gamma_m$ at large $m$. Expanding around $t=0$,
\begin{align}
    (t-T_c)^{-5} + (t+T_c)^{-5} = -\frac{2}{T_c^5} \sum_{m\geq1} \binom{2m+3}{4} \Big(\frac{t}{T_c} \Big)^{2m-1} ,
\end{align}
and the terms with $\bar T_c$ give the complex conjugate. So \eqref{eq:fourpoles} contributes
\begin{align}
    -4C_c \binom{2m+3}{4} {\rm Re} \big[T_c^{-2m-4} \big]
\end{align}
to the coefficient of $t^{2m-1}$. At large $m$, the coefficients of a power series are given by the sum of such contributions from its singularities. Therefore,
\begin{align} \label{eq:gammalarge}
    \gamma_m\simeq-\frac32\binom{2m+3}{4}\sum_{c=1,2}A_c\,{\rm Re}\Big[T_c^{-2m-4}\big(1+O(m^{-1})\big)\Big] .
\end{align}
The $O(m^{-1})$ corrections come from the $1/\omega$ corrections to the tail, which give the weaker singularities $(t-T_c)^{-4}, \, (t-T_c)^{-3}, \,  \cdots$.

\begin{figure}[H]
    \centering
    \includegraphics[width=0.7\textwidth]{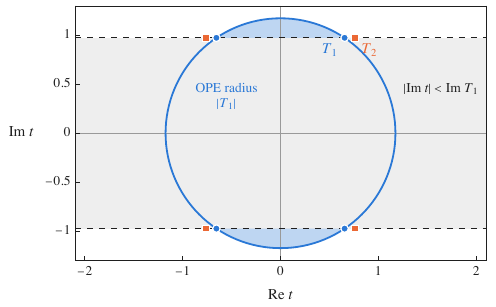}
    \caption{The complex $t$-plane for $(3,1,\tfrac15)$. The integral \eqref{eq:GRsine} converges in the gray strip $|{\rm Im}\,t|<{\rm Im}\,T_1$, and the spectral tail gives singularities on its edges, at $\pm T_1$, $\pm\bar T_1$ and $\pm T_2$, $\pm\bar T_2$. The blue circle is $|t|=|T_1|$, and the blue parts of the disk lie outside the strip.}
    \label{fig:OPEdisk}
\end{figure}

Since $T_1=T_2-B$, $T_1$ is closer to $t=0$ than $T_2$, so we expect the radius of convergence to be $|T_1|$. The spectral tail alone does not prove this. The integral \eqref{eq:GRsine} converges only inside the strip, but since $|T_1|>{\rm Im}\,T_1$, the disk $|t|<|T_1|$ also contains points outside the strip, near the imaginary axis, shown in blue in Fig.~\ref{fig:OPEdisk}. The spectral tail does not tell us whether $G_R(t)$ is singular there. If it were, the radius of convergence would be smaller than $|T_1|$. The coefficients $\gamma_m$, on the other hand, are controlled at large $m$ by the singularity or singularities nearest to $t=0$, regardless of their direction in the complex $t$-plane. We therefore locate the nearest singularity directly from $\gamma_m$, which we have computed up to $m=397$.

We define
\begin{align}
    s_m\equiv\frac{\gamma_m}{\binom{2m+3}{4}} .
\end{align}
At large $m$, the nearest singularity dominates. We call it $T_c$, without assuming which point it is or what its amplitude is. With $z_c=T_c^{-2}$, $s_m$ is then a linear combination of $z_c^{\,m+2}$ and $\bar z_c^{\,m+2}$ with constant coefficients, up to the $O(m^{-1})$ corrections. Since $z_c$ and $\bar z_c$ are the roots of $x^2- a x+ b$ with $a=z_c+\bar z_c$ and $b=|z_c|^2$, the sequence $s_m$ satisfies
\begin{align} \label{eq:threeterm}
    s_{m+1} = a  \, s_m- b\, s_{m-1} .
\end{align}
We treat $a$ and $b$ as unknowns. Imposing \eqref{eq:threeterm} at two consecutive values of $m$ fixes $a,b$ from four consecutive coefficients, and the nearest singularity follows from
\begin{align}
    |T_c| = b^{-1/4} , \qquad \cos(2\arg T_c)=\frac{a}{2\sqrt{b}} , \qquad 0<\arg T_c<\frac{\pi}{2} .
\end{align}
Because of the $O(m^{-1})$ corrections, \eqref{eq:threeterm} holds only approximately. Even so, at $m=395$ the estimate agrees with $T_1$ within $2\times10^{-6}$ for $(3,1,\tfrac15)$ and $5\times10^{-8}$ for $(2,\tfrac12,0)$, as shown in Table~\ref{tab:blindT1}. The difference is much smaller than the distance $B=0.105$ and $0.167$ between $T_1$ and $T_2$. The optical times $T_{1,2,3}$ do not enter this estimate, so $\gamma_m$ alone locates the nearest singularity at $T_1$.

\begin{table}[H]
\centering
\begin{tabular}{lllll}
\hline
 & & ${\rm Re}\,T_c$ & ${\rm Im}\,T_c$ & $|\Delta T_c|$ \\
\hline
$(3,1,\tfrac15)$ & $T_1$ (optical time) & $0.6550170488$ & $0.9748769088$ & \\
 & $T_c$ from $\gamma_m$ & $0.6550154704$ & $0.9748762461$ & $2\times10^{-6}$ \\
$(2,\tfrac12,0)$ & $T_1$ (optical time) & $0.7916069411$ & $1.1947240304$ & \\
 & $T_c$ from $\gamma_m$ & $0.7916069925$ & $1.1947240467$ & $5\times10^{-8}$ \\
\hline
\end{tabular}
\caption{The nearest singularity of $G_R(t)$, located from $\gamma_m$ alone by \eqref{eq:threeterm} at $m=395$, compared with the optical time $T_1$. $|\Delta T_c|$ is the distance between the two. For comparison, $T_2=0.75970+0.97488\,i$ for $(3,1,\tfrac15)$ and $T_2=0.95882+1.19472\,i$ for $(2,\tfrac12,0)$.}
\label{tab:blindT1}
\end{table}

Conversely, we fix $T_1$ and $T_2$ to the optical times and test the amplitudes $A_c$ in \eqref{eq:gammalarge}. Fitting $A_1$, $A_2$ and the $1/m$ corrections, we obtain $A_1=-2$ within $2\times10^{-14}$ and $A_2=-2$ within $2\times10^{-3}$ in both examples.

Therefore, the OPE convergence radius is controlled by $T_1$. It is $|T_1|=1.1745$ for $(3,1,\tfrac15)$ and $1.4332$ for $(2,\tfrac12,0)$, smaller than $|T_3|=1.3030$ and $1.7642$ of the physical singularity.

\section{QNM lines}
\label{sec:QNM}

The QNMs are the solutions of $\mathcal{A}(\omega)=0$. If $\omega$ is a QNM, so is $-\bar\omega$, so we consider ${\rm Re}\,\omega\geq0$ and write $\omega=|\omega|e^{-i\theta}$ with $0<\theta\leq\pi/2$.

For this $\omega$, $e^{i\omega(T-T_c)}$ is real and larger than $1$ on the ray $T=T_c-ire^{i\theta}$ with $r>0$. This is the Stokes line of $T_c$, which is the vertical line of Sec.~\ref{sec:reflect} tilted to the right by the angle $\theta$, as in Fig.~\ref{fig:Stokes}(b). For $0<\theta<\pi/2$, the Stokes lines of $T_1$, $T_2$ and $T_3$ all reach the real $T$ axis. In particular, the Stokes line of $T_3$ no longer runs along the vertical edge, so $T_3$ is not shadowed. The calculation of Sec.~\ref{sec:reflect} then gives
\begin{align} \label{eq:Acomplex}
    \mathcal{A}\propto1-\tilde\rho=1+2\sum_{c=1}^{3}\cos(\pi\nu_c)\,e^{i\omega T_c} ,
\end{align}
from \eqref{eq:rhotilde} and \eqref{1pmtilderho}. Since
\begin{align}
    \big|e^{i\omega T_c}\big|  &= \big| \exp \left(  i |\omega| e^{-i\theta} \, |T_c| e^{i \arg  T_c} \right)    \big|  \\
    &=  \exp \left(  -|\omega| \, |T_c|\sin(\arg T_c-\theta)  \right)  \nn  ,
\end{align}
$e^{i\omega T_c}$ is exponentially small for $\theta<\arg T_c$ and exponentially large for $\theta>\arg T_c$. Also, it has absolute value $1$ at $\theta=\arg T_c$.

So, as $\theta$ grows from $0$, $\mathcal{A}=0$ first becomes possible at the smallest $\arg T_c$. Since $T_1=T_2-B$ and $T_3=T_2+iG$, this is always $\arg T_2$. At $\theta=\arg T_2$, the terms of $T_1$ and $T_3$ are exponentially small and the term of $T_2$ has absolute value $1$. With $\nu_2=\frac13$, $\mathcal{A}=0$ becomes $1+e^{i\omega T_2}=0$, so
\begin{align} \label{eq:tower}
    \omega\simeq\frac{(2k+1)\pi}{T_2}
\end{align}
for large integers $k$.
This is a line of QNMs along the direction of $1/T_2$, that is, at $\arg\omega=-\arg T_2$, where $\arg T_2$ is the angle of the red line in Fig.~\ref{fig:Time}. For $\arg T_2<\theta<\pi/2$, the term of $T_2$ in \eqref{eq:Acomplex} is exponentially larger than all the other terms, including $1$, so we expect no QNM there. In particular, there is no QNM line along $1/T_1$ or $1/T_3$.

We compute the QNMs with $|\omega|<125$ as the zeros of $J(\omega)=\frac{\mathcal{A}(\omega)}{\Gamma(1-i\omega\beta/2\pi)}$ of \eqref{eq:J(w)}, following the method explained in Appendix~\ref{sec:QNMcal}. Fig.~\ref{fig:QNM} shows the numerical results. The complex QNMs approach the direction $-\arg T_2$, which is $-52.07^\circ$ for $(3,1,\tfrac15)$ and $-51.25^\circ$ for $(2,\tfrac12,0)$. The QNM with the largest $|\omega|$ in Fig.~\ref{fig:QNM} is $\omega=76.5170-98.2487i$ for $(3,1,\tfrac15)$ and $\omega=75.7098-94.3654i$ for $(2,\tfrac12,0)$, which give $\omega T_2/\pi=48.991-0.014i$ and $58.993-0.009i$. These are close to the odd integers $49$ and $59$, that is, to \eqref{eq:tower} with $k=24$ and $k=29$. For these QNMs, the distance $|\omega T_2/\pi-(2k+1)|$ is $0.017$ and $0.011$. For the complex QNMs with $|\omega|>30$, this distance decreases as $|\omega|$ grows.

To check that no QNM is missed, we count the zeros of $J$ in a region $\Omega$ of the complex $\omega$-plane with the argument principle,
\begin{align}
    N_\Omega = \frac{1}{2\pi i}\oint_{\partial\Omega}\frac{J'(\omega)}{J(\omega)}\,d\omega .
\end{align}
In general, $N_\Omega$ is the number of zeros of $J$ in $\Omega$ minus the number of poles of $J$ in $\Omega$. Since $J$ has no poles, $N_\Omega$ is the number of zeros of $J$ in $\Omega$, that is, the number of QNMs in $\Omega$. We take regions $r_1<|\omega|<r_2$, $\alpha_1<\arg\omega<\alpha_2$, which together cover the disk $|\omega|<125$.

\begin{figure}[H]
    \centering
    \includegraphics[width=\textwidth]{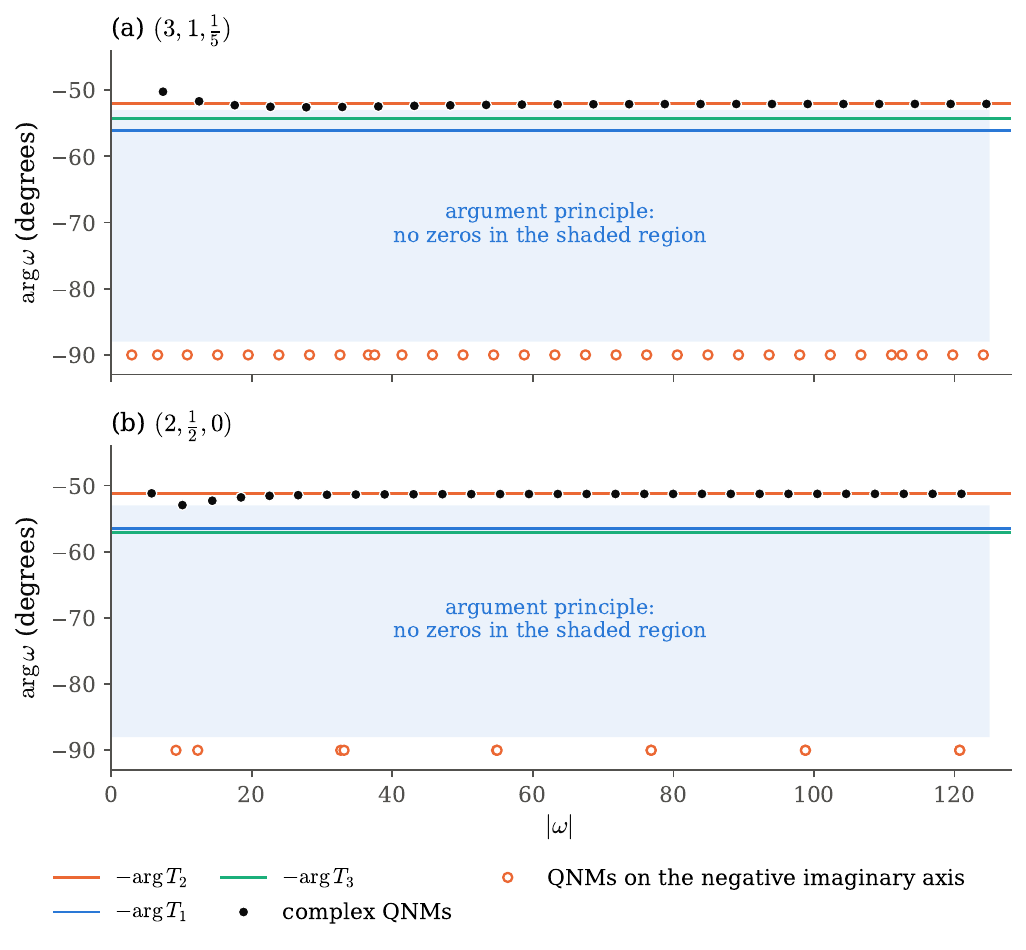}
    \caption{$\arg\omega$ against $|\omega|$ for the QNMs with ${\rm Re}\,\omega\geq0$. The filled circles are the complex QNMs and the open circles are the QNMs on the negative imaginary axis. The horizontal lines are the directions $-\arg T_c$. There is no zero of $J(\omega)$ in the shaded regions.}
    \label{fig:QNM}
\end{figure}

For $(3,1,\tfrac15)$, there is no zero in $-88^\circ<\arg\omega<-53^\circ$, which contains $-\arg T_1=-56.10^\circ$ and $-\arg T_3=-54.33^\circ$. The region $-53^\circ<\arg\omega<-49.5^\circ$ around $-\arg T_2$ contains 24 zeros, and $-92^\circ<\arg\omega<-88^\circ$ contains 31 zeros.

For $(2,\tfrac12,0)$, there is again no zero in $-88^\circ<\arg\omega<-53^\circ$, which contains $-\arg T_1=-56.47^\circ$ and $-\arg T_3=-57.08^\circ$. The region $-53^\circ<\arg\omega<-48^\circ$ around $-\arg T_2$ contains 29 zeros, and $-92^\circ < \arg \omega < -88^\circ$ contains 12 zeros.

In both examples, there is also no zero between the line of QNMs and the positive real axis, and no zero in the upper half-plane. Each count equals the number of QNMs that we found in the same region, so no QNM is missed.

We now turn to the QNMs on the negative imaginary axis, $\omega=-iy$ with $y>0$, shown as open red circles in Fig.~\ref{fig:QNM}. For $\omega=-iy$, the term of $T_3$ in \eqref{eq:Acomplex} has the same size as the term of $T_2$, since $|e^{i\omega(T_3-T_2)}|=|e^{-\omega G}|=1$. The calculation of Sec.~\ref{sec:reflect} does not apply here. There, $\phi_{\rm in}$ is followed from the horizon to the boundary along the real $T$ axis, which is the image of $R>1$, and a reflected wave is added at each point where the real $T$ axis crosses a Stokes line. For $\theta=\pi/2$, the Stokes line of each $T_c$ is the horizontal ray $T=T_c+r$ with $r>0$, as in Fig.~\ref{fig:Stokes}(c), and none of these rays crosses the real $T$ axis, so $\phi_{\rm in}$ picks up no reflected wave. Instead, we use the fact that $\mathcal{A}=0$ means $\phi_{\rm in}\propto\phi_B$, and compare the two solutions on the line ${\rm Im}\,T=\beta/2$, which is the image of $R_-<R<1$. Since the wave equation \eqref{eq:waveR} has real coefficients for $\omega=-iy$, $\phi_{\rm in}$ and $\phi_B$ can both be taken real for $R>1$, so $\phi_{\rm in}\propto\phi_B$ can hold only with a real constant. Two solutions that are proportional for $R>1$ stay proportional, with the same constant, for all $R$ and $T$. Therefore, on the line ${\rm Im}\,T=\beta/2$, $\phi_B$ must have the same phase as $\phi_{\rm in}$.

We first find the phase of $\phi_{\rm in}$. Near the horizon, $\phi_{\rm in}=\epsilon^s(1+a_1\epsilon+\cdots)$ with $\epsilon=R-1$, as in Appendix~\ref{sec:A(w)cal}. For $\omega=-iy$, the exponent $s=-\frac{i\omega\beta}{4\pi}=-\frac{y\beta}{4\pi}$ and the coefficients $a_n$ are real, so $\phi_{\rm in}$ is real for $R>1$. To reach $R<1$, we pass $R=1$ through the lower half $R$-plane, as for the optical times. Then $\epsilon=|\epsilon|e^{-i\pi}$, and $\epsilon^s = |\epsilon|^s e^{iy\beta/4}$ acquires a phase. Since the equation is real, this phase stays the same for all $R_-<R<1$, so on the line ${\rm Im}\,T=\beta/2$, i.e., for $R_-<R<1$, we have
\begin{align} \label{eq:phaseinf}
    \phi_{\rm in}=e^{iy\beta/4}\times({\rm real}) .
\end{align}

Next, we bring $\phi_B$ to the line ${\rm Im}\,T=\beta/2$. For $\omega=-iy$, the WKB waves are $f_\pm=P^{-1/4}e^{\pm yT/2}$. Near the boundary, $\phi_B\propto z^2I_2(yz)$, which grows inward, so $\phi_B\propto f_+$ on the real $T$ axis, since $T$ increases as $R$ decreases from the boundary. We follow it along the path of Fig.~\ref{fig:Stokes}(c): along the real $T$ axis to $T=p$, and then straight up to $T=p+i\beta/2$. We take $p>{\rm Re}\,T_2$, so that the path crosses none of the black lines in Fig.~\ref{fig:Stokes}(c), which are the image of the real $R$ axis. The path then reaches $R_-<R<1$ through the lower half $R$-plane, as for $\phi_{\rm in}$ in \eqref{eq:phaseinf}. On the way up, the path crosses the Stokes lines of $T_2$ and $T_3$. Each crossing adds a reflected wave as in \eqref{eq:reflect}, with the opposite sign since the path goes around $T_c$ in the opposite direction:
\begin{align} \label{eq:phiBline}
    \phi_B\propto f_++2i\sum_{c=2,3}\cos(\pi\nu_c)\,e^{yT_c}\,f_- .
\end{align}

On the line ${\rm Im}\,T =  \beta/2$, $T={\rm Re}\,T+i\beta/2$ and $P>0$, so $f_\pm=e^{\pm iy\beta/4}\times({\rm real})$. $f_+$ already acquires the same phase as \eqref{eq:phaseinf}, and hence $\phi_{\rm in} \propto \phi_B$ requires the second term of \eqref{eq:phiBline} to have the same phase. Since $f_-$ acquires the opposite phase $e^{-iy\beta/4}$, its coefficient must be $e^{iy\beta/2}$ times a real number. Since ${\rm Re}\,T_2={\rm Re}\,T_3$ and
\begin{align}
    e^{yT_c} &  = e^{y \, {\rm Re}\, T_2} \,  e^{iy \, {\rm Im }T_c} \\
    &= e^{y\,{\rm Re}\,T_2} \,e^{iy\beta/2} \, e^{-ih_cy} , \qquad (c=2,3) ,  \nn
\end{align}
with $h_c := \frac{\beta}{2} - {\rm Im}\,T_c$, we have to require
\begin{align}
    i \sum_{c=2,3} \cos(\pi \nu_c)\,  e^{-ih_cy} \in \RR,
\end{align}
that is,
\begin{align} \label{eq:E}
    E(y) := \sum_{c=2,3}\cos(\pi\nu_c)\cos(h_cy)=0 .
\end{align}

For $q_3>0$, both $R=-q_2$ and $R=-q_3$ are simple zeros of $P$, so $\nu_2=\nu_3=\frac13$ and we obtain
\begin{align}
    E(y)\propto\cos h_2y+\cos h_3y=2\cos\frac{(h_2+h_3)y}{2}\,\cos\frac{Gy}{2},
\end{align}
where we used $h_2-h_3 = {\rm Im} \left( T_3-T_2 \right) =G$. This gives two families of QNMs on the negative imaginary axis,
\begin{align} \label{eq:twofamilies}
    y\simeq\frac{(2k+1)\pi}{h_2+h_3} , \qquad y\simeq\frac{(2k+1)\pi}{G} .
\end{align}

For $q_3=0$, $T_3$ is the Kasner singularity. Near $R=0$, both $P\propto R$ and $D\propto R$ vanish, so $x$ diverges logarithmically and $P$ vanishes exponentially in $x$. The local solutions are Bessel functions of order zero, so $\nu_3=0$. Moreover, $\beta_-=0$ gives $h_3=0$ and $h_2=G$, so $T_3$ lies on the line ${\rm Im}\,T=\beta/2$ itself. For $\omega=-iy$, the Stokes line of $T_3$ is the horizontal ray to its right, so our path ends on the Stokes line of $T_3$ instead of crossing it, as in Fig.~\ref{fig:Stokes}(d). On the Stokes line itself, the Stokes multiplier of the reflected component is one half of its fully switched-on value \cite{Barry:1989zz}, so the $c=3$ term of \eqref{eq:E} has an extra factor $\frac12$. With $\cos(\pi\nu_2)=\frac12$ and $\cos(\pi\nu_3)=1$, we obtain
\begin{align} \label{eq:doubleE}
    E(y)\propto\cos Gy+1=2\cos^2\frac{Gy}{2} ,
\end{align}
which has double zeros at $y=(2k+1)\pi/G$.

We now include the effect of $T_1$ and determine how these double zeros split. For $\omega = - iy $, the Stokes line of $T_1$ runs along the horizontal edge into $T_2$, so $T_1$ is shadowed, as $T_3$ is for real $\omega$. Since $\omega(T_1-T_2)=iyB$ has the same form as $\omega(T_3-T_2)=i\omega G$ for real $\omega$ of Sec.~\ref{sec:tail}, the term of $T_2$ is multiplied by $W(yB)$ as in \eqref{eq:tailW}. Using $|\Gamma(\frac12-\frac{ig}{2\pi})|^2=\pi/\cosh(\frac g2)$, we have $|W(g)|^2=1+e^{-g}$ from \eqref{eq:defofW}, and
\begin{align}
    E(y)\propto1+|W(yB)|\cos(Gy+\arg W) ,
\end{align}
with $|W(yB)|=\sqrt{1+e^{-yB}}$. The phase of $W$ shifts the center of each pair of zeros. At large $y$, its variation is subleading, while the splitting is controlled at leading order by the excess $|W|-1$. The minima of $\cos Gy+1$ only touch zero, and the slightly larger amplitude $|W|>1$ of the cosine pushes them below zero. Near $y_0=(2k+1)\pi/G$, we have $\cos Gy\simeq-1+\frac12G^2(y-y_0)^2$ and $|W|\simeq1+\frac12e^{-By_0}$, so
\begin{align}
    E\propto-\frac12e^{-By_0}+\frac12G^2(y-y_0)^2 ,
\end{align}
and each double zero splits into two simple zeros separated by 
\begin{align} \label{eq:split}
    \Delta y\simeq\frac{2}{G}\,e^{-By/2} .
\end{align}

\begin{figure}[H]
    \centering
    \includegraphics[width=\textwidth]{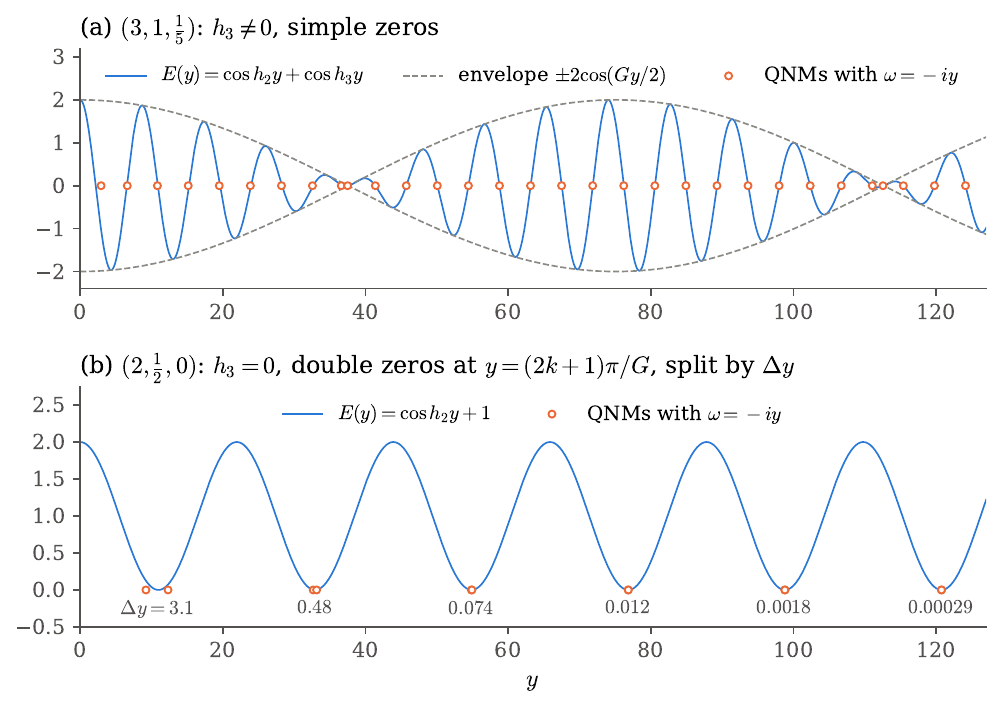}
    \caption{QNMs on the negative imaginary axis, $\omega=-iy$, compared with $E(y)$. (a) $(3,1,\tfrac15)$, with the envelope $\pm2\cos(Gy/2)$. (b) $(2,\tfrac12,0)$, with the splitting $\Delta y$ of each pair.}
    \label{fig:ImQNM}
\end{figure}

Fig.~\ref{fig:ImQNM} compares the QNMs on the negative imaginary axis with the zeros of $E(y)$. For $(3,1,\tfrac15)$, where $h_2=0.7633$ and $h_3=0.6796$, the 31 QNMs with $y<125$ correspond one to one to the 31 zeros of $E(y)$. For the first family of \eqref{eq:twofamilies}, the difference is below $0.013$ for $y>40$. The second family is set by the distance $G$ between $T_2$ and $T_3$. It predicts $y=37.54$ and $112.61$, and the QNMs are at $y=37.49$ and $112.53$.

For $(2,\tfrac12,0)$, the 12 QNMs form six pairs, one around each double zero $y=(2k+1)\pi/G$, which are $y=10.98$, $32.93$, $54.88$, $76.83$, $98.78$ and $120.73$. The pairs are split by $\Delta y=3.10$, $0.475$, $0.0738$, $0.0117$, $0.00185$ and $0.000294$. These are $1.11$, $1.07$, $1.04$, $1.03$, $1.02$ and $1.02$ times \eqref{eq:split}, so the ratio approaches $1$ as $y$ grows.

Therefore, the complex QNMs are controlled by $T_2$, and the QNMs on the negative imaginary axis are controlled by $T_2$ and $T_3$. The physical singularity $T_3$ appears at leading order only in the latter. $T_1$ has no QNM line of its own. For $q_3=0$, it appears instead through the exponentially small splitting of the double zeros on the negative imaginary axis.

\section{Conclusion and discussions}
\label{sec:conclusion}

We studied the retarded correlator of $T_{xy}$ at zero spatial momentum in the STU black branes with unequal R-charges. Their metric has three singular points at $R=-q_i$, and only $R=-q_3$ is the physical singularity. The optical times form a right triangle with $T_1=T_2-B$ and $T_3=T_2+iG$, and the three observables detect different points of this triangle.

The spectral function at large real $\omega$ is controlled by $T_1$ and $T_2$, which have the smallest ${\rm Im}\,T_i$. Their two reflection terms have the same amplitude $A_1=A_2=-2$ in \eqref{eq:tailLO}, and they beat with the period $2\pi/B$. The physical singularity $T_3$ is shadowed by $T_2$, and it contributes only through the additional factor $W(\omega G)$ of the $T_2$ reflection term. For $\omega G\gg1$, this factor amounts to a reflection term of $T_3$ with half of its amplitude, the value on a Stokes line.

The OPE converges up to $|T_1|$, the smallest $|T_i|$, which is smaller than $|T_3|$ of the physical singularity. The OPE coefficients alone locate the nearest singularity at $T_1$, without using the optical times. Although $R=-q_1$ lies beyond the physical singularity, $T_1$ is on the principal sheet of $G_R(t)$.

The complex QNMs lie along the direction of $1/T_2$, since $T_2$ has the smallest $\arg T_i$. The QNMs on the negative imaginary axis $\omega=-iy$ are set by $T_2$ and $T_3$, so the physical singularity appears at leading order only there. For $q_3=0$, the physical singularity is a Kasner singularity, and its optical time $T_3$ lies on the line ${\rm Im}\,T=\beta/2$. The QNMs on the negative imaginary axis then come in pairs near the double zeros of $E(y)$ \eqref{eq:doubleE}, and $T_1$ splits each pair by an exponentially small amount $\Delta y$ \eqref{eq:split}.

Conversely, the spectral function, the OPE coefficients, and the QNMs are all data of the dual CFT. From them, one can read off the positions of several singular points of the geometry, including those beyond the physical singularity, which are not part of the spacetime. One can even read off the order of each zero of $P$. The amplitude $A_c=-4\cos(\pi\nu_c)$ with $\nu_c=1/(\ell+2)$ fixes the order $\ell$, and our fits of the spectral function and of the OPE coefficients recover $A_1=A_2=-2$, that is, simple zeros. The QNM line carries the same information. If $R=-q_2$ were a zero of another order, the condition $1+2\cos(\pi\nu_2)\,e^{i\omega T_2}=0$ would give $\omega T_2=(2k+1) \pi + i \ln \big( 2 \cos(\pi\nu_2) \big)$, so the offset of $\omega T_2/\pi$ from the odd integers measures $\nu_2$. For a simple zero, $2\cos(\pi\nu_2)=1$ and the offset vanishes, consistent with Sec.~\ref{sec:QNM}.

There are several interesting open directions. The singular points beyond the physical singularity are not part of the spacetime, but they control the boundary data. It would be interesting to understand what they correspond to in the dual field theory, and how their signatures change at finite $N$, where the bouncing singularities are smoothed out \cite{Komargodski:2026oob, Batra:2026gbq, Dodelson:2026hjm, Liu:2026luf}. The charged black holes in AdS$_4$ and AdS$_7$ carry four and two charges, respectively \cite{Cvetic:1999xp}, and their metrics also have several singular points, so the same question can be asked there. Finally, other channels, such as the R-current correlators, involve coupled fluctuations at finite density, and they may detect the singular points differently.

\vskip 0.5cm

\hspace*{-0.8cm} {\bf\large Acknowledgements}
\vskip 0.2cm

\hspace*{-0.75cm}
I thank Seok Kim for helpful discussions. I also thank Joydeep Chakravarty for valuable comments on the subleading contribution of the physical singularity to the spectral function. This work is supported in part by the NRF grant RS-2026-25477643. I used Claude Code for numerical calculations and figures.

\appendix

\section{Computing $\mathcal{A}(\omega)$}
\label{sec:A(w)cal}

In this section, we explain how to compute $\mathcal{A}(\omega)$ for a given $\omega$. We need it for the spectral function through \eqref{eq:rhoA}, and the QNMs are its zeros. The difficulty is that $\phi_{\rm in}$ is defined by its behavior at the horizon, while $\mathcal{A}$ is read off at the boundary, and both are singular points of the wave equation.

\subsection{Wronskian}

For two solutions $\phi_1,\phi_2$ of the wave equation \eqref{eq:waveR}, we define the Wronskian
\begin{align}
    W[\phi_1, \phi_2] = \phi_1 \, \partial_R \phi_2 - \phi_2 \, \partial_R \phi_1 .
\end{align}
Then, $D W$ is independent of $R$:
\begin{align}
    \partial_R (DW) &= \phi_1 \, \partial_R (D \, \partial_R \phi_2) - \phi_2 \, \partial_R (D \, \partial_R \phi_1) \\
    &= -\frac{\omega^2 P}{4D} \left( \phi_1 \phi_2 - \phi_2 \phi_1 \right) = 0 . \nn
\end{align}
At the boundary, the solutions $\phi_A$ and $\phi_B$ of Sec.~\ref{sec:obs} behave as $\phi_A \simeq 1$ and $\phi_B \simeq R^{-2}$, and $D\simeq R^3$, so $W[\phi_A,\phi_B]\simeq-2R^{-3}$ and
\begin{align} \label{eq:DW}
    D\,W[\phi_A,\phi_B] = -2 .
\end{align}
Using $\phi_{\rm in} = \mathcal{A} \phi_A + \mathcal{B} \phi_B$ and $W[\phi_B, \phi_B]=0$, we have
\begin{align}
    W[\phi_{\rm in}, \phi_B] = \mathcal{A}\, W[\phi_A,\phi_B] = - \frac{2}{D} \mathcal{A} ,
\end{align}
so we can find $\mathcal{A}(\omega)$ as
\begin{align} \label{eq:A(w)}
    \mathcal{A}(\omega) = -\frac{D}{2} \left( \phi_{\rm in} \, \partial_R \phi_B - \phi_B \, \partial_R \phi_{\rm in} \right) .
\end{align}
This can be computed at any $R$. Therefore, we only need to know $\phi_{\rm in}$, $\phi_B$ and their derivatives, at the same point $R$.

\subsection{Frobenius expansions}

The wave equation \eqref{eq:waveR} is
\begin{align}   \label{eq:waveinR}
    D^2 \partial_R^2 \phi + D \, (\partial_R D) \, \partial_R \phi + \frac{\omega^2 P}{4} \phi =0   ,
\end{align}
and dividing this by $D^2$, we can write
\begin{align}
    \partial_R^2 \phi + \frac{\partial_R D}{D} \partial_R  \phi + \frac{\omega^2 P}{4 D^2} \phi =0 .
\end{align}
At finite $R$, the coefficients diverge only at the zeros of $D$, that is at $R=1,R_-,R_3$. There, the coefficient of $\partial_R\phi$ has at most a simple pole and the coefficient of $\phi$ has at most a double pole, so these are regular singular points. The boundary $R=\infty$ is also a regular singular point, as we see from \eqref{waveinu} below. Therefore, we can apply the Frobenius expansion around each of them. An expansion around a singular point converges up to the nearest other singular point. For example, the expansion around the outer horizon $R=1$ converges in the region $|R-1|<1-R_-$.

We need the expansions around the horizon and the boundary, which give $\phi_{\rm in}$ and $\phi_B$. These are singular points of the wave equation, so a numerical integration cannot start there. The infalling condition at the horizon and the normalizable condition at the boundary are statements about the behavior at these points. The Frobenius expansions turn them into values of $\phi$ and $\partial_R\phi$ at regular points, with any desired precision. When the convergence regions of the two expansions do not overlap, we connect them by solving the wave equation numerically.

We start with the horizon. With $\epsilon= R-1$, we can write
\begin{align}
    D = d_1 \epsilon + d_2 \epsilon^2 + d_3 \epsilon^3, \qquad P = p_0+ p_1 \epsilon + p_2 \epsilon^2 + p_3 \epsilon^3. 
\end{align}

At $\epsilon \to 0$, the leading order of the wave equation \eqref{eq:waveinR} is
\begin{align}
    (d_1\epsilon)^2 \cdot  \partial_\epsilon^2 \phi + d_1^2 \epsilon \cdot  \partial_\epsilon \phi  + \frac{\omega^2 p_0}{4}  \cdot \phi   = 0.
\end{align}

If we insert $\phi= \epsilon^s$,
\begin{align}
    d_1^2 \cdot s(s-1) + d_1^2 \cdot s + \frac{\omega^2 p_0 }{4} =0 \ \to \ d_1^2 \cdot s^2 + \frac{\omega^2 p_0}{4} =0 ,
\end{align}
so we obtain
\begin{align}
    s= \pm \frac{i\omega\sqrt{p_0}}{2d_1} = \pm \frac{i \omega \beta}{4\pi},
\end{align}
where we used $d_1 = D'(1),  \, p_0 = P(1)$ and $\beta = \frac{2\pi \sqrt{P(1)}}{D'(1)}$. We have to find which one is the infalling solution. Near the horizon, the tortoise coordinate $S$ of Sec.~\ref{sec:reflect} satisfies
\begin{align}
    dS = -\frac{\sqrt{P}}{2D}\,dR \simeq -\frac{\sqrt{p_0}}{2d_1\epsilon}\,d\epsilon = -\frac{\beta}{4\pi}\frac{d\epsilon}{\epsilon} ,
\end{align}
and
\begin{align}
    S \simeq -\frac{\beta}{4\pi}\ln\epsilon \ \to \ +\infty .
\end{align}
The infalling wave is $e^{-i\omega(t-S)}$ and
\begin{align}
    \phi \propto e^{i\omega S} = \epsilon^{-i\omega\beta/4\pi} .
\end{align}
Therefore, $s=-\frac{i\omega\beta}{4\pi}$ corresponds to the infalling wave.

We set the Frobenius expansion:
\begin{align}
    \phi_{\rm in} = \sum_{n \geq 0} a_n \epsilon^{n+s}, \qquad a_0=1. 
\end{align}

For brevity, we write
\begin{align}
    D^2 = \sum_{k=2}^6 e_k \epsilon^k, \quad D \, (\partial_RD) = \sum_{k=1}^5 g_k \epsilon^k, \quad P = \sum_{k=0}^3 p_k \epsilon^k . 
\end{align}

Each term of \eqref{eq:waveinR} becomes
\begin{align}
    & D^2 \, \partial_R^2\phi =  \sum_{k,n} e_k a_n (n+s)(n+s-1) \epsilon^{n+s+k-2} ,  \\
    & D \, (\partial_R D) \,  \partial_R \phi =  \sum_{k,n} g_k a_n (n+s)\epsilon^{n+s+k-1} ,   \nn \\
    & \frac{\omega^2 P}{4} \phi = \frac{\omega^2}{4} \sum_{k,n} p_k a_n \epsilon^{n+s+k} . \nn
\end{align}
By collecting the coefficient of $\epsilon^{N+s}$, we obtain
\begin{align} \label{eq:a_n}
    \sum_{n=0}^N a_n \left[  (n+s) (n+s-1)  e_{N-n+2} + (n+s)  g_{N-n+1} +  \frac{\omega^2 }{4} p_{N-n}  \right] =0 .
\end{align}
When $n=N$ in the summation, the coefficient of $a_n$ is
\begin{align}
    & (N+s)(N+s-1)e_2 + (N+s)g_1 + \frac{\omega^2 p_0}{4} \\
    &= (N+s)^2 d_1^2 + \frac{\omega^2 p_0}{4}     \nn \\
    &= (N+s)^2 d_1^2 -d_1^2 s^2 \nn \\
    &= d_1^2 \,  N(N+2s) ,  \nn
\end{align}
where we used $e_2 =  g_1 =d_1^2$ and $\omega^2 p_0=-4d_1^2 s^2$. Therefore, when $N=0$, \eqref{eq:a_n} automatically holds. This is the freedom of the overall constant factor and we have already fixed it by $a_0=1$. For $N>0$, we can find $a_N$ recursively from \eqref{eq:a_n}:
\begin{align}  \label{eq:a_N}
    a_N &= -\frac{1}{d_1^2 \, N( N+2s) } \sum_{n=0}^{N-1} a_n  \left[ (n+s) (n+s-1)  e_{N-n+2} +  (n+s)  g_{N-n+1} +  \frac{\omega^2}{4} p_{N-n}  \right] . 
\end{align}
As we already mentioned, this expansion converges only for $|\epsilon| < 1-R_-$. Note that the denominator becomes zero when $N+2s=0$. Since $s= - \frac{i\beta \omega}{4\pi}$, it means
\begin{align}
    \omega = \frac{2\pi N}{ i \beta  } .
\end{align}
These are the Matsubara frequencies in the lower half-plane. The infalling solution $\phi_{\rm in}$ has poles at these frequencies, and they also appear in $\mathcal{A}(\omega)$. We can remove these poles from $\mathcal{A}(\omega)$ by dividing it by $\Gamma(1+2s) = \Gamma \left( 1- i\omega \beta/ 2\pi   \right)$, since $\Gamma(n) $ has poles at $n=0,-1,-2,\cdots $.

Next, we consider the boundary. With $u=1/R$, we define
\begin{align}
    H(u) \equiv H_1 H_2 H_3 = \prod_i (1+q_i u)= u^3 P, \qquad F(u) \equiv H(u) - M u^2 = u^3 D .
\end{align}
Using $\partial_R = -u^2 \partial_u$, we obtain
\begin{align}
    & D\,  \partial_R \phi = \frac{F}{u^3} (-u^2) \partial_u \phi  = -\frac{F}{u} \partial_u \phi , \\
    & \partial_R (D \,  \partial_R \phi) = u^2 \partial_u \left[  \frac{F}{u} \partial_u \phi  \right] = uF \phi_{uu} +(uF_u-F) \phi_u   , \nn
\end{align}
where a subscript $u$ denotes $\partial_u$. The wave equation $\partial_R(D \, \partial_R \phi) = -\frac{\omega^2 P}{4D} \phi$ becomes, after multiplying by $F$,
\begin{align}  \label{waveinu}
    uF^2 \phi_{uu} + (uF F_u - F^2) \phi_u + \frac{\omega^2 H}{4} \phi =0 .
\end{align}

Since $F(0)=H(0)=1$ and $F_u(0) = \sum_i q_i$, the leading equation is
\begin{align}
    u\phi_{uu} - \phi_u =0 .
\end{align}
By inserting $\phi= u^\alpha$, we obtain
\begin{align}
    \alpha (\alpha-1) - \alpha = \alpha(\alpha-2) =  0. 
\end{align}
So, $\alpha=0$ or $\alpha=2$. Near the boundary, a scalar dual to an operator of dimension $\Delta$ in $d$ dimensions has the non-normalizable mode $r^{\Delta-d}$, which is the source, and the normalizable mode $r^{-\Delta}$, which is the response. For our massless $\phi$, $\Delta=d=4$, so these are $r^0=1$ and $r^{-4}=R^{-2}=u^2$, which correspond to $\alpha=0$ and $\alpha=2$, respectively.

So, $\phi_A$ is the non-normalizable solution and $\phi_B$ is the normalizable one. We only need $\phi_B$, and we write it as
\begin{align}
    \phi_B = \sum_{n \geq 0} b_n u^{n+2}, \quad b_0=1.
\end{align}
Just as in the near-horizon case, we write
\begin{align}
    F^2 = \sum_{k=0}^6 e_k u^k, \quad F F_u = \sum_{k=0}^5 g_k u^k, \quad H = \sum_{k=0}^3 p_k u^k . 
\end{align}
The next steps are almost the same as near the horizon. By collecting the coefficient of $u^{N+1}$, we obtain
\begin{align} \label{eq:b_n}
    & \sum_{n=0}^N b_n \left[  (n+2) (n+1)  e_{N-n} +   (n+2) \left(   g_{N-n-1}  -e_{N-n} \right) +  \frac{ \omega^2}{4} p_{N-n-1}  \right] \\
    &=  \sum_{n=0}^N b_n \left[  (n+2) n \,  e_{N-n} +   (n+2)  g_{N-n-1}  +  \frac{ \omega^2}{4} p_{N-n-1}  \right]     =0 .   \nn 
\end{align}
When $n=N$, using $e_0 = 1, \,  g_{-1} = p_{-1} =0$, we find that the coefficient of $b_n$ in the summation is $(N+2)N$. Therefore, when $N=0$, \eqref{eq:b_n} automatically holds. Just as near the horizon, this is the freedom of the overall constant factor and we have already fixed it by $b_0=1$. For $N>0$, we can find $b_N$ recursively from \eqref{eq:b_n}:
\begin{align}   \label{eq:b_N}
    b_N &= -\frac{1}{ (N+2)N } \sum_{n=0}^{N-1} b_n  \left[  (n+2) n \,   e_{N-n} +  (n+2)  g_{N-n-1} +  \frac{\omega^2}{4} p_{N-n-1}  \right] . 
\end{align}
For example, $b_1= - \frac13 \left[ 2 g_0 + \frac{\omega^2}{4} p_0   \right] $ where $g_0= q_1+q_2+q_3 , \, p_0=1$. This series converges in the region $|u|<\min \left( 1, \frac{1}{|R_3|} \right) $, since the singular points are at $u=1$, $u=1/R_3$ and $u=1/R_->1$.

\subsection{Spectral function in terms of $\mathcal{A}$}

We now derive \eqref{eq:rhoA}. We do not need to calculate $\mathcal{B}$. For real $\omega$, $\phi_{\rm in}^*$ is also a solution, so $D\,W[\phi_{\rm in}, \phi_{\rm in}^*]$ is independent of $R$. Near the horizon, $\phi_{\rm in}\simeq\epsilon^{s}$ and $\phi_{\rm in}^*\simeq\epsilon^{-s}$ because $s=-\frac{i\omega \beta}{4\pi}$, so
\begin{align}
    D\,W[\phi_{\rm in}, \phi_{\rm in}^*] \simeq d_1 \epsilon \cdot \left( -\frac{2s}{\epsilon} \right) = \frac{i\omega\beta d_1}{2\pi} = i\omega\sqrt{P(1)} .
\end{align}
On the other hand, $\phi_A$ and $\phi_B$ are real for real $\omega$, so $\phi_{\rm in}^* = \mathcal{A}^* \phi_A + \mathcal{B}^* \phi_B$ and
\begin{align}
    D\,W[\phi_{\rm in}, \phi_{\rm in}^*] = \left( \mathcal{A} \mathcal{B}^* - \mathcal{A}^* \mathcal{B} \right) D\,W[\phi_A, \phi_B] = 4i \, {\rm Im} \left( \mathcal{A}^* \mathcal{B} \right) ,
\end{align}
where we used \eqref{eq:DW}. Therefore, ${\rm Im}\left( \mathcal{A}^*\mathcal{B} \right) = \omega\sqrt{P(1)}/4$, and
\begin{align}
    \rho = {\rm Im}\,\frac{\mathcal{B}}{\mathcal{A}} = \frac{{\rm Im}\left( \mathcal{A}^*\mathcal{B} \right)}{|\mathcal{A}|^2} = \frac{\omega\sqrt{P(1)}}{4|\mathcal{A}|^2} .
\end{align}
This formula holds for the normalization $|\phi_{\rm in}|=1$ at the horizon, which follows from $a_0=1$.

Next, we calculate $\rho_{\rm vac}$. In the vacuum, we have $q_i=0$ and $M=0$, so $P=D=R^3$ and \eqref{eq:waveR} becomes
\begin{align}
    \partial_R \left( R^3 \, \partial_R \phi \right) = -\frac{\omega^2}{4} \phi .
\end{align}
With $z=1/\sqrt{R}$, its solutions are $z^2 J_2(\omega z)$ and $z^2 Y_2(\omega z)$, or equivalently $z^2 H^{(1,2)}_2(\omega z)$. Using
\begin{align}
    J_2(\zeta) \simeq \frac{\zeta^2}{8} , \qquad Y_2(\zeta) \simeq -\frac{4}{\pi \zeta^2} , \qquad (\zeta\to 0) ,
\end{align}
we have $z^2 J_2(\omega z)\simeq\frac{\omega^2}{8} z^4 = \frac{\omega^2}{8} R^{-2}$ and $z^2 Y_2(\omega z)\simeq-\frac{4}{\pi\omega^2}$ near the boundary. So,
\begin{align}
    \phi_B = \frac{8}{\omega^2}\, z^2 J_2(\omega z) , \qquad
    \phi_A = -\frac{\pi\omega^2}{4}\, z^2 Y_2(\omega z) .
\end{align}
In the vacuum, the tortoise coordinate is $S=z$, and $z^2 H^{(1)}_2(\omega z)\propto z^{3/2} e^{i\omega z}$ at large $\omega z$, so this is the infalling solution. Its overall factor drops out of $\mathcal{B} / \mathcal{A}$, so we may take
\begin{align}
    \phi_{\rm in} = z^2 H^{(1)}_2(\omega z) = z^2 J_2(\omega z) + i z^2  Y_2(\omega z) = -\frac{4i}{\pi\omega^2}\,\phi_A + \frac{\omega^2}{8}\,\phi_B .
\end{align}
Therefore, we obtain
\begin{align}
    \mathcal{A}_{\rm vac} = -\frac{4i}{\pi\omega^2} , \qquad \mathcal{B}_{\rm vac} = \frac{\omega^2}{8} , \qquad
    \rho_{\rm vac} = {\rm Im}\,\frac{\mathcal{B}_{\rm vac}}{\mathcal{A}_{\rm vac}} = \frac{\pi\omega^4}{32} .
\end{align}

\subsection{Numerics and QNMs}
\label{sec:QNMcal}

\begin{figure}[H]
    \centering
    \captionsetup[subfigure]{singlelinecheck=false, justification=raggedright, skip=2pt}
    \begin{subfigure}{\textwidth}
        \caption{}
        \includegraphics[width=\textwidth]{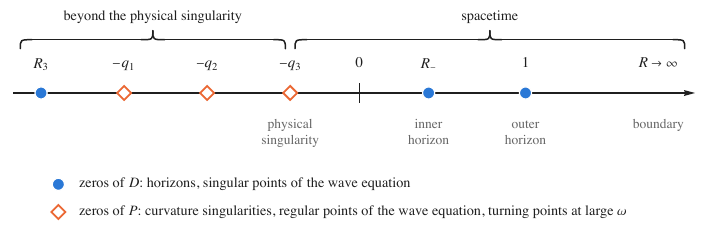}
    \end{subfigure}
    \begin{subfigure}{\textwidth}
        \caption{}
        \includegraphics[width=\textwidth]{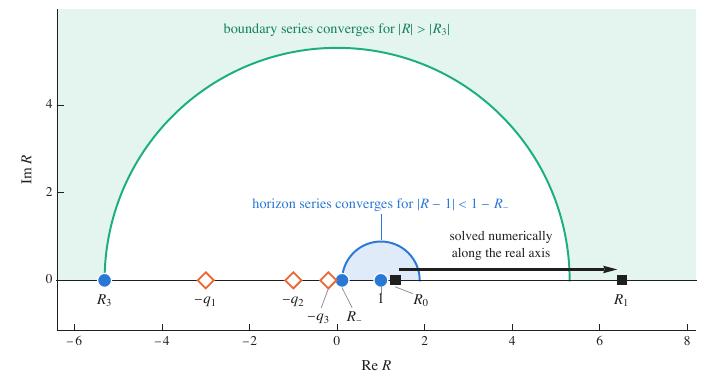}
    \end{subfigure}
    \caption{(a) The order of the zeros of $D$ and of $P$ on the real $R$ axis. (b) The upper half of the complex $R$-plane for $(q_1,q_2,q_3)=\left( 3,1,\tfrac15 \right)$. The regions of convergence of the horizon series and of the boundary series do not overlap, so we carry the solution from $R_0$ to $R_1$ along the real axis.}
    \label{fig:Match}
\end{figure}

For a given $\omega$, we choose a point $R_0$ inside the region of convergence of the Frobenius series near the horizon, and evaluate $\phi_{\rm in}$ and $\partial_R\phi_{\rm in}$ there from the series. We then solve the wave equation numerically from $R_0$ to a point $R_1$, where $R_1$ is inside the region of convergence of the series of $\phi_B$ near the boundary, as shown in Fig.~\ref{fig:Match}.

Using the values of $\phi_{\rm in}, \partial_R \phi_{\rm in}, \phi_B, \partial_R \phi_B$ at $R_1$, the formula \eqref{eq:A(w)} gives $\mathcal{A}(\omega)$. From this $\mathcal{A}(\omega)$, we can obtain $\rho(\omega)$ using \eqref{eq:rhoA}. In Sec.~\ref{sec:tail}, we compare $e^{\omega\,{\rm Im}\,T_1} \, \Delta\rho/\rho_{\rm vac}$ with the WKB expectation. Since $e^{\omega\,{\rm Im}\,T_1}$ is very large, we need $\Delta\rho/\rho_{\rm vac}$ to high accuracy: at $\omega=120$, it is as small as $10^{-50}$ for $\left( 3,1,\tfrac15 \right)$ and $10^{-62}$ for $\left( 2,\tfrac12,0 \right)$. We therefore use 100 digits for real $\omega$ in $16\leq\omega\leq120$.

The QNMs are the solutions of $\mathcal{A}(\omega)=0$. To remove the Matsubara poles of $\mathcal{A}(\omega)$, we define
\begin{align} \label{eq:J(w)}
    J(\omega) = \frac{\mathcal{A}(\omega)}{\Gamma\left(1-i\omega\beta/2\pi\right)} ,
\end{align}
which has no poles. We evaluate $J(\omega)$ in the same way, and find its zeros with the secant method,
\begin{align}
    \omega_{k+1} = \omega_k - J(\omega_k)\,\frac{\omega_k-\omega_{k-1}}{J(\omega_k)-J(\omega_{k-1})} .
\end{align}
$J$ also vanishes at a Matsubara frequency if the sum in \eqref{eq:a_N} vanishes there, since then $\mathcal{A}$ stays finite. We checked in exact arithmetic that this does not happen for $N\leq6$ in our two examples. We compute each zero with $0.7|\omega|+25$ digits.

\section{OPE coefficients}
\label{sec:OPEcal}

In this section, we show that, for $t>0$ and in the normalization $G_R(\omega)=\mathcal{B}/\mathcal{A}$,
\begin{align} \label{eq:GR(t)}
    G_R(t) = \frac32 \, t^{-5} + \sum_{m\geq1} \gamma_m \, t^{2m-1} ,
\end{align}
for the metric \eqref{eq:STUmetric}, and explain how to compute $\gamma_m$ from the large-$\omega$ expansion of $G_R(\omega) $. Unlike $\mathcal{A}(\omega)$ in Appendix~\ref{sec:A(w)cal}, this does not require solving the wave equation from the horizon to the boundary. We solve the wave equation only around the boundary, order by order in $1/\omega^2$. We then read off $G_R(\omega)$, and transform it to $G_R(t)$.

\subsection{Near-boundary expansion in $1/\omega^2$}

Dividing \eqref{waveinu} by $F^2$ gives
\begin{align}
    u\phi_{uu}+\Big(\frac{uF_u}{F}-1\Big)\phi_u+\frac{\omega^2}{4}\frac{H}{F^2}\,\phi=0 .
\end{align}

The vacuum solutions of Appendix~\ref{sec:A(w)cal} depend on $u$ through $\omega^2 u$, so at large $\omega$ we introduce
\begin{align}
    v=-\omega^2u .
\end{align}

With $\partial_u=-\omega^2\partial_v$ and after dividing by $-\omega^2$, the equation becomes
\begin{align} \label{eq:inner}
    L_0\phi := v\phi_{vv}-\phi_v-\frac14\phi=Q_1(u)\,\phi_v+Q_2(u)\,\phi , \qquad u=-\frac{v}{\omega^2} ,
\end{align}
where
\begin{align}
    Q_1(u)=-\frac{uF_u}{F}=\sum_{k\geq1}Q_{1,k}\,u^k , \qquad
    Q_2(u)=\frac14\Big(\frac{H}{F^2}-1\Big)=\sum_{k\geq1}Q_{2,k}\,u^k .
\end{align}
Both $Q_1$ and $Q_2$ vanish at $u=0$ because $F(0)=H(0)=1$. The operator $L_0$ does not contain $\omega$, and $\omega$ enters only through $u=-v/\omega^2$ in $Q_1$ and $Q_2$.

In the vacuum, $Q_1=Q_2=0$, and the two solutions of $L_0\phi=0$ are $vK_2(\sqrt v)$ and $vI_2(\sqrt v)$. The one that decays into the bulk for ${\rm Im}\,\omega>0$ is
\begin{align}
    \chi(v)=v\,K_2\big(\sqrt v\big) , \qquad \sqrt v=-i\omega\sqrt u ,
\end{align}
which is proportional to the vacuum infalling solution $z^2H^{(1)}_2(\omega z)$ of Appendix~\ref{sec:A(w)cal}, with $z=\sqrt u$.

Since $u=-v/\omega^2$, the right-hand side of \eqref{eq:inner} is a power series in $\omega^{-2}$,
\begin{align}
    Q_1\phi_v+Q_2\phi=\sum_{k\geq1}\omega^{-2k}(-v)^k\big(Q_{1,k}\,\phi_v+Q_{2,k}\,\phi\big) .
\end{align}
We therefore expand
\begin{align}
    \phi=\sum_{n\geq0}\omega^{-2n}\phi_n , \qquad \phi_0=\chi (v) ,
\end{align}
and solve, order by order,
\begin{align} \label{eq:Lphin}
    L_0\phi_n=\sum_{k=1}^{n}(-v)^k\big(Q_{1,k}\,\partial_v\phi_{n-k}+Q_{2,k}\,\phi_{n-k}\big) .
\end{align}
We look for each order in the form
\begin{align}
    \phi_n=\lambda_n(v)\,\chi+\mu_n(v)\,\chi' ,
\end{align}
with polynomials $\lambda_n$ and $\mu_n$, where $'=d/dv$. This form closes under $\partial_v$ and $L_0$, because $\chi''$ is a combination of $\chi$ and $\chi'$. Since $\phi_0= \chi$, we have $\lambda_0=1$ and $\mu_0=0$.

From $L_0\chi=0$, we have $\chi''=(\chi'+\chi/4)/v$. Hence, for any polynomials $\lambda$ and $\mu$, we obtain
\begin{align} \label{eq:dchi}
    \big (\lambda\chi+\mu\chi'\big)' &=    \lambda' \chi + (  \lambda+\mu') \chi' + \mu \chi''   \\
    &= \lambda' \chi + (\lambda +\mu') \chi'+ \frac{\mu}{v} \left( \chi' + \frac{\chi}{4} \right)   \nn \\
    &=  \Big(\lambda'+\frac{\mu}{4v}\Big)\chi+\Big(\lambda+\mu'+\frac{\mu}{v}\Big)\chi' , \nn
\end{align}
and therefore,
\begin{align} \label{eq:L0lm}
    & L_0\big[\lambda\chi+\mu\chi'\big] \\
    &=  v ( \lambda \chi + \mu \chi')'' - (\lambda \chi +\mu \chi')'  - \frac14 (\lambda \chi+ \mu \chi')   \nn \\
    &=   v \left[  \Big(\lambda'+\frac{\mu}{4v}\Big)\chi+\Big(\lambda+\mu'+\frac{\mu}{v}\Big)\chi'  \right]'   -   \left[  \Big(\lambda'+\frac{\mu}{4v}\Big)\chi+\Big(\lambda+\mu'+\frac{\mu}{v}\Big)\chi'  \right] - \frac14 (\lambda \chi+ \mu \chi')     \nn \\
    &= v \left[ \left(   \lambda'' + \frac{\mu'}{4v} - \frac{\mu}{4v^2}  \right) \chi  + \left( \lambda' + \frac{\mu}{4v} +\lambda' + \mu'' + \frac{\mu'}{v} - \frac{\mu}{v^2}  \right) \chi' \right.  \nn \\
    &\qquad \qquad \left. +  \left( \lambda+ \mu' + \frac{\mu}{v}  \right) \frac{1}{v} \left(  \chi' + \frac{\chi}{4} \right) \right]    -   \left[  \Big(\lambda'+\frac{\mu}{4v}\Big)\chi+\Big(\lambda+\mu'+\frac{\mu}{v}\Big)\chi'  \right] - \frac14 (\lambda \chi+ \mu \chi')         \nn \\
    & = \Big(v\lambda''-\lambda'+\frac{\mu'}{2}-\frac{\mu}{4v}\Big)\chi
    +\Big(2v\lambda'+v\mu''+\mu'-\frac{\mu}{v}\Big)\chi'  . \nn
\end{align}
All coefficients in \eqref{eq:dchi} and \eqref{eq:L0lm} are polynomials only if $\mu(0)=0$, so we require $\mu_n(0)=0$ for all $n$.

The polynomial $\lambda_n$ has one more freedom. Equation \eqref{eq:Lphin} fixes $\phi_n$ only up to a solution of $L_0\phi=0$, and only $\chi$ is infalling, so $\phi_n$ is fixed up to $c_n\chi$ with a constant $c_n$. This corresponds to multiplying $\phi$ by an overall factor that depends on $\omega$, which cancels in $\mathcal{B}/\mathcal{A}$. We use this freedom to set $\lambda_n(0)=0$ for $n\geq1$.

At order $n$, the right-hand side of \eqref{eq:Lphin} contains only $\phi_0,\ldots,\phi_{n-1}$, which are already known. Using \eqref{eq:dchi} for $\partial_v\phi_{n-k}$, it takes the form $S_n(v)\chi+T_n(v)\chi'$ with the known polynomials
\begin{align} \label{eq:ST_n}
    S_n&=\sum_{k=1}^{n}(-v)^k\Big[Q_{1,k}\Big(\lambda_{n-k}'+\frac{\mu_{n-k}}{4v}\Big)+Q_{2,k}\,\lambda_{n-k}\Big] , \\
    T_n&=\sum_{k=1}^{n}(-v)^k\Big[Q_{1,k}\Big(\lambda_{n-k}+\mu_{n-k}'+\frac{\mu_{n-k}}{v}\Big)+Q_{2,k}\,\mu_{n-k}\Big] . \nn
\end{align}
For instance, $\lambda_0=1$ and $\mu_0=0$ give
\begin{align} \label{eq:ST_1}
    S_1 = -Q_{2,1}v , \qquad T_1 =- Q_{1,1} v .
\end{align}
Since $\chi$ and $\chi'$ are independent, \eqref{eq:Lphin} and \eqref{eq:L0lm} give
\begin{align}
    v\lambda_n''-\lambda_n'+\frac{\mu_n'}{2}-\frac{\mu_n}{4v}=S_n , \qquad
    2v\lambda_n'+v\mu_n''+\mu_n'-\frac{\mu_n}{v}=T_n .
\end{align}
We write
\begin{align}
    \lambda_n=\sum_{d \geq 0 }\lambda_{n,d}v^d , \qquad \mu_n=\sum_{d \geq 0 } \mu_{n,d}v^d , \qquad S_n=\sum_{d \geq 0 }S_{n,d}v^d , \qquad T_n=\sum_{d \geq 0 } T_{n,d}v^d ,
\end{align}
where $\lambda_{0,0}=1$, $\lambda_{n,0}=0$ for $n\geq1$, $\mu_{n,0}=0$ for all $n$, and $S_{n,0}=T_{n,0}=0$ since every term of $S_n$ and $T_n$ carries a factor $v^k$ with $k\geq1$. The coefficients of $v^d$ give
\begin{align}
    &(d+1)d  \, \lambda_{n, d+1} - (d+1) \,  \lambda_{n,d+1} +\frac{(d+1)}{2} \, \mu_{n, d+1} - \frac{1}{4}  \, \mu_{n, d+1} = S_{n,d} , \\
    & 2 d \, \lambda_{n,d} +(d+1)d \, \mu_{n, d+1}  + (d+1) \, \mu_{n,d+1} - \mu_{n,d+1} = T_{n,d} , \nn
\end{align}
that is,
\begin{align} \label{eq:coefmatch}
    (d+1)(d-1)\,\lambda_{n,d+1}+\frac{2d+1}{4}\,\mu_{n,d+1}&=S_{n,d} , \\
    2d\,\lambda_{n,d}+d(d+2)\,\mu_{n,d+1}&=T_{n,d} . \nn
\end{align}

When we know $S_n,T_n$, we can find $\lambda_{n,d}$ and $\mu_{n,d}$ for all $d$, as follows. The first equation contains $\lambda_{n,d+1}$ and $\mu_{n,d+1}$, and the second contains $\lambda_{n,d}$ and $\mu_{n,d+1}$. Let $v^K$ be the highest power appearing in $S_n$ or $T_n$. We take $\lambda_{n,d}=0$ for $d>K$ and $\mu_{n,d}=0$ for $d>K+1$, and solve one unknown at a time starting from the highest $d$. The first equation at $d=K$ gives $\mu_{n,K+1}$, the second at $d=K$ then gives $\lambda_{n,K}$, the first at $d=K-1$ gives $\mu_{n,K}$, and so on down to $\mu_{n,1}$. At $d=0$, the second equation reads $0=T_{n,0}$, which holds automatically. The first equation at $d=0$ gives
\begin{align} \label{eq:mu1}
    \mu_{n,1}=4\lambda_{n,1} .
\end{align}
This determines $\lambda_n$ and $\mu_n$ from $S_n$ and $T_n$.

Starting from $\lambda_0=1$ and $\mu_0=0$, we can obtain $S_1,T_1$ using \eqref{eq:ST_n}, which gives \eqref{eq:ST_1}. From $S_1,T_1$, we find $\lambda_1, \mu_1$ using the process above. From $\lambda_0,\mu_0,\lambda_1, \mu_1$, we can obtain $S_2,T_2$ using \eqref{eq:ST_n}. We repeat this order by order: $\lambda_n,\cdots, \lambda_0$ and $\mu_n,\cdots, \mu_0$ give $S_{n+1}$ and $T_{n+1}$, which give $\lambda_{n+1}$ and $\mu_{n+1}$.

\subsection{Reading off $G_R(\omega)$}

Near $v=0$, $\chi(v) = v\, K_2(\sqrt{v})$ has the series expansion
\begin{align}
    \chi(v) =  2-\frac v2+\frac{v^2}{8}\Big(k_E-\frac12\log v\Big)+O\big(v^3\log v\big) , \qquad k_E=\log2-\gamma_E+\frac34 ,
\end{align}
where $\gamma_E$ is Euler's constant. Hence,
\begin{align}
    \phi_n &= \lambda_n \chi + \mu_n \chi' \\
    &=  \left( \lambda_{n,0} + \lambda_{n,1} v + \lambda_{n,2} v^2 + O(v^3)  \right) \cdot \left[   2- \frac{v}{2} + \frac{v^2}{8} \left( k_E- \frac12 \log v  \right) + O(v^3 \log v) \right]    \nn \\
    &\qquad   + \left(  \mu_{n,1}v + \mu_{n,2} v^2 + O(v^3) \right) \cdot \left[  -\frac12 + \frac{v}{4} \left(  k_E - \frac14   \right) -\frac{v}{8} \log v + O(v^2 \log v )   \right]  \nn  \\
    &= 2 \lambda_{n,0} +v \left( 2\lambda_{n,1} -\frac{\lambda_{n,0}}{2} -\frac{\mu_{n,1}}{2}   \right)  \nn \\
    &\qquad   + v^2 \left( 2 \lambda_{n,2} -\frac{\lambda_{n,1}}{2} + \frac{\lambda_{n,0}}{8} k_E  -\frac{\mu_{n,2}}{2}  +\frac{\mu_{n,1}}{4} \left( k_E- \frac14 \right)  \right) \nn \\
    &\qquad   + v^2 \log v \left( -\frac{\lambda_{n,0}}{16} - \frac{\mu_{n,1}}{8} \right) + O(v^3 \log v) . \nn
\end{align}
Since $\lambda_{n,0} = \delta_{n,0}$, the coefficient of $v^0$ is $2\delta_{n,0}$, and by \eqref{eq:mu1} the coefficient of $v$ is $-\frac12 \delta_{n,0}$. We define $X_n$ and $\tilde{X}_n$ as the coefficients of $v^2$ and $v^2\log v$:
\begin{align}
    X_n   &\equiv   \frac{k_E}{8}\lambda_{n,0}-\frac12\lambda_{n,1}+2\lambda_{n,2}+\Big(\frac{k_E}{4}-\frac1{16}\Big)\mu_{n,1}-\frac12\mu_{n,2} , \\
    \tilde X_n   &\equiv  -\frac1{16}\lambda_{n,0}-\frac18\mu_{n,1} . \nn
\end{align}
Summing over $n$, we obtain
\begin{align}
    \phi = 2 -\frac12 v  + v^2 \sum_{n\geq0} \omega^{-2n} X_n + v^2  \log v \sum_{n\geq0} \omega^{-2n} \tilde X_n + O(v^3\log v) .
\end{align}
With $v=-\omega^2u$ and $\log v=\log(-\omega^2)+\log u$, this becomes
\begin{align}
    \phi = 2 + \frac12 \omega^2  u + \omega^4 u^2 \sum_{n\geq0} \omega^{-2n} \Big[ X_n + \tilde X_n \log(-\omega^2) \Big]  + \omega^4 u^2 \log u \sum_{n\geq0} \omega^{-2n} \tilde X_n + O(u^3\log u) .
\end{align}

We compare this with $\phi = \mathcal{A} \phi_A + \mathcal{B} \phi_B$ of Sec.~\ref{sec:obs}, where $\phi_A\simeq1$ and $\phi_B\simeq u^2$. The $u^0$ coefficient gives $\mathcal{A}=2$. The terms in $u$ and $u^2\log u$ come only from $\phi_A$ and are fixed by $\mathcal{A}$. The $u^2$ coefficient is $\mathcal{B}+\mathcal{A}\,a_2$, where $a_2$ is the $u^2$ coefficient of $\phi_A$.

The condition $\phi_A\simeq1$ does not fix $\phi_A$ uniquely, since $\phi_A+c\,\phi_B$ also starts with $1$. This changes $a_2$ by $c$ and $\mathcal{B}$ by $-c\,\mathcal{A}$, so $G_R=\mathcal{B}/\mathcal{A}$ changes by $-c$. We choose $a_2$ to be a real polynomial in $\omega^2$, so that $G_R$ is fixed up to such polynomials. These are contact terms: since $\int dt\,e^{i\omega t}\,\delta^{(k)}(t)=(-i\omega)^k$, a polynomial in $\omega$ is the Fourier transform of a combination of $\delta(t)$ and its derivatives, which vanishes for $t>0$. A real polynomial also has no imaginary part on the real axis. Hence, up to contact terms,
\begin{align} \label{eq:GRomega}
    G_R(\omega)=\frac{\mathcal{B}}{\mathcal{A}}=\frac{\omega^4}{2}\sum_{n\geq0}\omega^{-2n}\Big[X_n+\tilde X_n\log(-\omega^2)\Big] .
\end{align}
Here $X_n$ and $\tilde X_n$ are independent of $\omega$. The $X_n$ terms with $n=0,1,2$ are polynomials in $\omega^2$, so they are also contact terms.

In the vacuum, $H=F=1$ and $Q_1=Q_2=0$. The right-hand side of \eqref{eq:Lphin} then vanishes at every order, and $\phi_n=0$ for $n\geq1$, so the vacuum correlator is the $n=0$ term of \eqref{eq:GRomega} alone. In the black brane, the $n=0$ term is the same, and the terms with $n\geq1$ are the corrections due to the background. For $n=0$, we have $\lambda_{0,0}=1$ and $\mu_{0,1}=0$, hence $\tilde X_0=-\frac1{16}$, and \eqref{eq:GRomega} contains $-\frac{\omega^4}{32}\log(-\omega^2)$. For real $\omega>0$, approached from ${\rm Im}\,\omega>0$, we have $\log(-\omega^2)=\log\omega^2-i\pi$, so ${\rm Im}\,G_R=\pi\omega^4/32$, which is $\rho_{\rm vac}$ of Sec.~\ref{sec:obs}.

We now prove that $\tilde X_n=0$ for all $n\geq1$ and any charges. For $n\geq1$, $\lambda_{n,0}=0$ and \eqref{eq:mu1} give
\begin{align} \label{eq:tildeX}
    \tilde X_n=-\frac18\,\mu_{n,1}=-\frac12\,\lambda_{n,1} .
\end{align}
At $d=1$, \eqref{eq:coefmatch} reads $\frac34\mu_{n,2}=S_{n,1}$ and $2\lambda_{n,1}+3\mu_{n,2}=T_{n,1}$, so
\begin{align} \label{lambda_n1}
    \lambda_{n,1}=\frac12\big(T_{n,1}-4S_{n,1}\big) .
\end{align}
The $k$-th term in $S_n$ and $T_n$ \eqref{eq:ST_n} carries the factor $(-v)^k$, so only $k=1$ contributes to $S_{n,1}$ and $T_{n,1}$:
\begin{align}
    S_{n,1}&=-Q_{1,1}\Big(\lambda_{n-1,1}+\frac14\mu_{n-1,1}\Big)-Q_{2,1}\,\lambda_{n-1,0} , \\
    T_{n,1}&=-Q_{1,1}\big(\lambda_{n-1,0}+2\mu_{n-1,1}\big) . \nn
\end{align}
Using \eqref{eq:mu1} at order $n-1$, the terms with $\lambda_{n-1,1}$ and $\mu_{n-1,1}$ cancel in $T_{n,1}-4S_{n,1}$, and \eqref{lambda_n1} becomes
\begin{align}
    \lambda_{n,1}=\frac12\big(4Q_{2,1}-Q_{1,1}\big)\lambda_{n-1,0} .
\end{align}
For $n\geq2$, this vanishes because $\lambda_{n-1,0}=0$. For $n=1$, we have $\lambda_{0,0}=1$, but $F(0)=H(0)=1$ and $F_u(0)=H_u(0)=\sum_iq_i$ give $Q_{1,1}=-\sum_iq_i$ and $Q_{2,1}=-\frac14\sum_iq_i$, so $4Q_{2,1}-Q_{1,1}=0$.

In conclusion, $\lambda_{n,1}=\mu_{n,1}=0$ and $\tilde X_n=0$ by \eqref{eq:tildeX} for all $n\geq1$. No logarithm appears beyond the vacuum one. The terms with $n=1,2$ are then only contact terms, and the terms with $n\geq3$ are negative powers of $\omega$, which give the OPE coefficients through \eqref{eq:gammaX} below.

\subsection{From $G_R(\omega)$ to $G_R(t)$}

Contact terms do not affect $G_R(t)$ at $t>0$. The remaining terms of \eqref{eq:GRomega} are fixed by the behavior of $G_R(t)$ near $t=0$: by Watson's lemma, the large-$\omega$ expansion of $G_R(\omega)=\int_0^\infty dt\,e^{i\omega t}G_R(t)$ is obtained by transforming the small-$t$ expansion of $G_R(t)$ term by term. For ${\rm Im}\,\omega>0$ and $m\geq1$,
\begin{align}
    \int_0^\infty dt\,e^{i\omega t}\,t^{2m-1}=(2m-1)!\,(-i\omega)^{-2m}=(-1)^m(2m-1)!\,\omega^{-2m} ,
\end{align}
so a term $\gamma_m\,t^{2m-1}$ of $G_R(t)$ gives $(-1)^m(2m-1)!\,\gamma_m\,\omega^{-2m}$. Since the term of \eqref{eq:GRomega} with $n=m+2$ is $\frac12X_{m+2}\,\omega^{-2m}$, we can obtain $\gamma_m$ from $X_{m+2}$ as
\begin{align} \label{eq:gammaX}
    \gamma_m=\frac{(-1)^m\,X_{m+2}}{2\,(2m-1)!} .
\end{align}

The integral of $t^{-5}$ diverges at $t=0$, so we shift the power of $t$ by a small $\varepsilon$:
\begin{align}
    \int_0^\infty dt\,e^{i\omega t}\,t^{-5+\varepsilon}
    =\Gamma(-4+\varepsilon)\,(-i\omega)^{4-\varepsilon}
    =\frac{\omega^4}{24\,\varepsilon}-\frac{\omega^4}{48}\log(-\omega^2)+{\rm const}\cdot\omega^4+O(\varepsilon) .
\end{align}
The divergent term and the term ${\rm const}\cdot\omega^4$ are polynomials in $\omega$, so they are contact terms. A term $c\,t^{-5}$ therefore gives $-\frac{c}{48}\,\omega^4\log(-\omega^2)$ up to contact terms, and the vacuum logarithm $-\frac{\omega^4}{32}\log(-\omega^2)$ requires $c=\frac32$. In the same way, $t^{-3}$ and $t^{-1}$ would give $\omega^2\log(-\omega^2)$ and $\log(-\omega^2)$, the logarithms of $n=1$ and $n=2$, and $t^{2m-1}\log t$ would give $\omega^{-2m}\log(-\omega^2)$, the logarithm of $n=m+2$. Since we proved $\tilde X_n=0$ for $n\geq1$, none of these terms appear. Even powers of $t$ would give odd powers of $\omega$, which do not appear either. Altogether, for $t>0$,
\begin{align}
    G_R(t)=\frac32\,t^{-5}+\sum_{m\geq1}\gamma_m\,t^{2m-1} .
\end{align}
For $\left( 3,1,\tfrac15 \right)$ the first orders are $X_3=-\tfrac{116352}{4375}$, $X_4=\tfrac{5973504}{21875}$ and $X_5=\tfrac{3523608576}{109375}$, which give $\gamma_1=\tfrac{58176}{4375}$, $\gamma_2=\tfrac{497792}{21875}$ and $\gamma_3=-\tfrac{73408512}{546875}$. For $\left( 2,\tfrac12,0 \right)$ they are $X_3=-\tfrac{458}{63}$, $X_4=\tfrac{5522}{105}$ and $X_5=\tfrac{897472}{231}$, which give $\gamma_1=\tfrac{229}{63}$, $\gamma_2=\tfrac{2761}{630}$ and $\gamma_3=-\tfrac{56092}{3465}$.

\bibliographystyle{JHEP}
\bibliography{refs}

\end{document}